\documentclass[12pt]{article}

\usepackage{amsmath,graphicx}
\usepackage{epstopdf}
\usepackage{multirow}
\usepackage{amsfonts}
\usepackage{amssymb}
\usepackage{amscd}
\usepackage{cite}

\def\hybrid{\topmargin -20pt    \oddsidemargin 0pt
        \headheight 0pt \headsep 0pt
        \textwidth 6.25in       
        \textheight 9.5in       
        \marginparwidth .875in
        \parskip 5pt plus 1pt   \jot = 1.5ex}

\hybrid

\numberwithin{equation}{section}
\numberwithin{table}{section}

\newcommand{\beq}{\begin{equation}\begin{aligned}}
\newcommand{\eeq}{\end{aligned}\end{equation}}
\newcommand{\bi}{\begin{itemize}}
\newcommand{\ei}{\end{itemize}}
\newcommand{\bea}{\begin{eqnarray}}
\newcommand{\eea}{\end{eqnarray}}
\newcommand{\ba}{\begin{array}}
\newcommand{\ea}{\end{array}}
\newcommand{\bt}{\begin{tabular}}
\newcommand{\et}{\end{tabular}}
\newcommand{\bc}{\begin{center}}
\newcommand{\ec}{\end{center}}

\newcommand{\cL}{\mathcal{L}}
\newcommand{\cK}{\mathcal{K}}
\newcommand{\cN}{\mathcal{N}}

\newcommand{\cF}{\mathcal{F}}

\newcommand{\cM}{\mathcal M}

\newcommand{\nn}{\nonumber}
\newcommand{\IM}{\textrm{Im} \,}
\newcommand{\RE}{\textrm{Re} \,}

\newcommand{\cref}{{\bf [check ref]}}

\newcommand{\M}{M}

\newcommand{\vfive}{\nu}

\begin{document}


\begin{titlepage}
\begin{center}


\rightline{\small WIS/11/08-JUN-DPP, ZMP-HH/08-8}
\vskip 1cm

{\Large \bf  Non-Abelian structures in compactifications of}
\vskip 0.3cm
{\Large \bf M-theory
on seven-manifolds with $SU(3)$ structure}

\vskip 1.2cm

{\bf Ofer Aharony$^{a}$, Micha Berkooz$^{a}$, Jan Louis$^{b,c}$ and
Andrei Micu$^{d}$\footnote{On leave from IFIN-HH Bucharest.} }
\
\vskip 0.8cm
{}$^{a}${\em Department of Particle Physics,\\ Weizmann Institute of Science,
Rehovot 76100, Israel}  \\
{\tt Ofer.Aharony@weizmann.ac.il, Micha.Berkooz@weizmann.ac.il}
\vskip 0.4cm

{}$^{b}${\em II. Institut f{\"u}r Theoretische Physik der Universit{\"a}t Hamburg\\
Luruper Chaussee 149,  D-22761 Hamburg, Germany}\\
 {\tt jan.louis@desy.de}
\vskip 0.4cm

{}$^{c}${\em Zentrum f\"ur Mathematische Physik,
Universit\"at Hamburg,\\
Bundesstrasse 55, D-20146 Hamburg}

\vskip 0.4cm

{}$^{d}${\em Physikalisches Institut der Universit\"at Bonn \\
  Nussallee 12, 53115 Bonn, Germany}\\
 {\tt amicu@th.physik.uni-bonn.de}
\end{center}

\vskip 1cm

\begin{center} {\bf ABSTRACT } \end{center}

\noindent We study  M-theory compactified on a specific class of
seven-dimensional manifolds with $SU(3)$ structure. The manifolds
can be viewed as a fibration of an arbitrary Calabi-Yau threefold
over a circle, with a U-duality twist around the circle. In some
cases we find that in the four dimensional low energy effective theory a (broken)
non-Abelian gauge group appears. Furthermore, such compactifications
are shown to be dual to previously analyzed compactifications of the
heterotic string on $K3\times T^2$, with background gauge field
fluxes on the $T^2$.

\vfill

June 2008

\end{titlepage}



\section{Introduction}


The study of possible space-time backgrounds of string theories has
been an active field of research for almost 25 years. A specific
subclass of
backgrounds admit a geometrical interpretation in which the space-time
manifold is the product space
\begin{equation}
M_d\, \times\, Y_{10-d}\ ,
\end{equation}
where $M_d$ is an infinitely extended $d$-dimensional manifold with
Minkowskian signature while $Y_{10-d}$ is a (10$-d$)-dimensional compact
manifold with Euclidean signature.  In standard compactifications, $Y$ is
constrained to be a Calabi-Yau manifold whose holonomy controls the amount of
unbroken supersymmetry present in the string background.  More generally, one
can turn on background fluxes for various $p$-form fields in the compact
directions, and then $Y$ is no longer constrained to be Ricci-flat.  Such
`generalized' compactifications have been studied intensively in recent years
\cite{grana}.

It has been observed early on that these generalized compactifications can be
discussed in terms of `manifolds with $G$-structure' \cite{waldram}. Such
manifolds admit a globally defined spinor (or tensor) which is left invariant
by the subgroup $G$ of the structure group. Generically such manifolds have
torsion and they can be characterized by a set of non-vanishing torsion
classes \cite{joyce,CS}. In string compactifications the number of invariant
spinors is directly related to the number of supersymmetries present in the
background. Calabi-Yau manifolds are a specific subclass of manifolds with
$G$-structure where the torsion vanishes and the invariant spinor is
covariantly constant with respect to the Levi-Civita connection.

String theories have the feature that their space-time
backgrounds can be dual to each other. This is firmly established for
dualities which hold in string perturbation theory.
For example type IIA string
theory compactified on a Calabi-Yau threefold $Y$ coincides with
type IIB string
theory compactified on the mirror Calabi-Yau $\tilde Y$.
For dualities which involve the dilaton (the string coupling) in a
non-trivial way, so far there is only (strong) evidence for the
validity of the duality. An example of such a duality is the heterotic
string compactified on $K3\times T^2$, which is believed to be
identical to type IIA string theory compactified
on a $K3$-fibred Calabi-Yau threefold \cite{K3d,KLM}.

An interesting question is the fate of these dualities for generalized
compactifications.  It has been shown in refs.~\cite{GLMW,GMPT,GLW,DFT} that mirror
symmetry continues to hold in the supergravity limit for compactifications on
manifolds with $SU(3)\times SU(3)$ structure. On the other hand, for the
heterotic--type II duality mentioned above only partial results have been
obtained so far \cite{CKKL,LM3}. More specifically it has been shown in
\cite{LM3} that in the supergravity limit a particular class of $SU(3)\times
SU(3)$-structure compactifications of type IIA is dual to the heterotic string
compactified on $K3\times T^2$ with a specific choice of background fluxes.
However, for some of the fluxes which can be turned on in the heterotic theory
no dual type IIA compactification could be identified. These are fluxes which
result in a gauged $N=2$ supergravity with vector multiplets carrying a
non-Abelian charge.

Let us review this in a little more detail. At the level of the
low-energy effective supergravity the heterotic -- type II duality
corresponds to a duality between $N=2$ supergravities in $d=4$.
Such supergravities can be coupled to $N=2$ vector-, hyper- and
tensor-multiplets.\footnote{A tensor multiplet can be dualized to a
  hypermultiplet or a vector multiplet, depending on the mass of the
  tensor.}
Turning on background fluxes or compactifying on manifolds with
$G$-structure correspond to the deformation of an ungauged supergravity into
a gauged or massive supergravity, with the flux and torsion playing the
role of non-trivial gauge charges or mass parameters.
For $SU(3)\times SU(3)$-structure compactifications of
type~II  it was shown in \cite{LM2,DDSV,GLMW,GLW,DFT,DFTV,BC}
that only gauged supergravities with charged hypermultiplets
or massive tensor multiplets appear. It was left as a puzzling feature
of such compactifications that charged vector multiplets could not be
obtained.
On the other hand in the heterotic $K3\times T^2$
compactification, non-Abelian gauge symmetries do appear precisely when
fluxes on the $T^2$ are turned on \cite{LM1}.

One of the goals of this paper is to resolve this puzzle.
We find that non-Abelian gauge symmetries also appear on the type IIA
side if instead of $SU(3)\times SU(3)$-structure compactifications of
type~II one considers compactifications of M-theory on 7-dimensional
manifolds with $SU(3)$ structure. We argue that these are the duals
of the heterotic compactifications with fluxes on the $T^2$. These are
the main results of our paper.

In this paper we do not attempt to work out the low-energy effective action
for a generic 7-dimensional $SU(3)$-structure manifold, but instead focus on a
very specific subclass of $SU(3)$-structure manifolds which lead to
non-Abelian gauge symmetries. More specifically, we consider 7-dimensional
manifolds which can be viewed as a non-trivial fibration of a Calabi-Yau
threefold $CY_3$ over a circle $S^1$. Furthermore, we impose that only the
second cohomology $H^{(1,1)}$ of $CY_3$ is twisted when going around the
$S^1$, but that the third cohomology $H^3$ (which governs the hypermultiplet
sector) is left unchanged. This constraint leads to a hypermultiplet sector
which is entirely determined by $H^3(CY_3)$ and therefore can be safely
neglected for the purpose of this paper.  It is precisely the twisted
$H^{(1,1)}$ cohomology which induces the non-Abelian structure into the
theory.  When $CY_3$ is $K3$-fibred and M-theory on $CY_3$ is dual to
heterotic string theory on $K3\times S^1$, twists of this type provide the
dual of the $T^2$ fluxes on the heterotic side.

Let us describe the twisting in slightly more detail. Consistency requires
that after going around the circle which takes us from 5 to 4 dimensions,
$H^{(1,1)}$ is rotated by an element of the U-duality group \cite{Hull,KM} of
the five-dimensional theory which corresponds to M-theory compactified
on $CY_3$ or the dual heterotic theory on $K3\times S^1$. In this case we
have $\Gamma({\bf Z}) = SO(1,h^{(1,1)}-2,{\bf Z})$, which on the heterotic side
is just the T-duality group.  On the M-theory side this symmetry exists
precisely for the dual K3-fibred Calabi-Yau manifolds \cite{KLM}.

In four dimensions the result is the appearance of non-Abelian gauge
symmetries as follows. The base of the K3-fibration is a ${\bf P_1}$
whose volume is identified with the four dimensional heterotic dilaton. Heterotic
weak coupling corresponds to a large ${\bf P_1}$-base,
and in that
limit the low-energy limits of both theories have a
$SO(2,h^{(1,1)}-1,{\bf R})$ symmetry in four dimensions. In this
case, we can describe the twisting in the language of four
dimensional supergravity as gauging\footnote{By gauging we mean that
isometries of the scalar manifold are mixed into the gauge
transformations, and not that new gauge fields are introduced.} an
isometry inside $SO(2,h^{(1,1)}-1, {\bf R})$. This makes the four-dimensional
gauge transformations non-commuting (non-Abelian).

At a generic point in field space, the non-Abelian gauge symmetry is
spontaneously broken, giving a mass to some of the gauge bosons.  We
find that on the M-theory side the gauge boson masses are inversely
proportional to the radius of the M-theory circle. In
order to consistently keep these gauge bosons in the low energy
effective action we need to require that these masses are smaller
than the Kaluza-Klein masses of the $CY_3$. This means that the
M-theory circle has to be larger than the radii of the Calabi-Yau,
which in turn forces us into the M-theory regime of type IIA string
theory. This is the reason that the non-Abelian structure is not
visible in $SU(3)\times SU(3)$ compactifications of type II
theories.

Let us stress that the flux on the heterotic side, and similarly the
non-trivial monodromy in the M-theory compactification, lead to a
non-trivial potential on the moduli space. We do not discuss here
the stabilization of these moduli, which can be accomplished by
adding additional ingredients. Instead, we just compute and compare
the resulting low-energy effective actions, without attempting to
solve their equations of motion.

The paper is organized as follows. In section~\ref{sec:geo} we discuss the
Kaluza-Klein (KK) reduction of M-theory on a seven-dimensional manifold with
$SU(3)$ structure. As a warm-up, we first recall in section~\ref{5dM} the
properties of the five-dimensional background corresponding to the reduction
of M-theory on a Calabi-Yau threefold. This sets the stage for the specific
$S^1$-fibration we consider in section~\ref{SU3}. In section~\ref{sec:KK} we
derive the low energy effective supergravity by a Kaluza-Klein reduction
from 11 to 4 dimensions, paying special attention to the gauging of the vector
multiplets. In section~\ref{sec:N2} we rewrite the effective action in a form
which shows the consistency with $N=2$ gauged supergravity. In
section~\ref{K3fibred} we consider the specific case of a $K3$-fibred
Calabi-Yau threefold which is the class of backgrounds dual to the heterotic
string. Most of our explicit computations are done in the limit in which the
scalar moduli space has a continuous isometry, as this makes the computations
simpler; in section~\ref{break} we discuss what happens when we go away from
this limit. In section~\ref{hetsect} we turn to the heterotic string
compactified on $K3\times T^2$ and start by recalling a few generic properties
of such backgrounds in section~\ref{genpro}. We then compare the mass-scales
in the dual backgrounds in section~\ref{mapmas}, showing the necessity to go
to the M-theory regime on the type II side when we turn on heterotic fluxes on
the $T^2$. In section~\ref{fluxmon} we argue that also the heterotic fluxes
can be viewed as a monodromy in the T-duality group. In section \ref{N2het} we
then recall the heterotic effective action as computed in \cite{LM1}. Finally,
in section~\ref{mcomparison} we compare the effective actions on both sides
and show that for a subset of torsion parameters they perfectly match. For the
convenience of the reader we briefly recall the vector multiplet sector of
(gauged) $N=2$ supergravity in appendix~\ref{sg}. Additional details of the
vector multiplets in heterotic string compactifications are assembled in
appendix~\ref{vshet}.


\section{M-theory compactifications on manifolds with $SU(3)$ structure}
\label{sec:geo}

In this section we compactify M-theory on seven-dimensional manifolds with
$SU(3)$ structure. By construction this leads to an $N=2$ supersymmetric
effective theory in $d=4$. However, as already explained in the introduction,
we do not consider the most general manifolds with $SU(3)$ structure but
instead focus on a particular subclass of manifolds which lead to a low-energy
supergravity with non-Abelian vector multiplets. For simplicity we further
insist that the moduli space of the hypermultiplets coincides with that of a
$CY_3 \times S^1$ compactification, where all scalars in hypermultiplets are
gauge neutral.  Thus, we do not pay attention to the hypermultiplets but only
concentrate on the vector multiplet sector.  The gaugings which appear in the
hypermultiplet sector in general compactifications of M-theory on manifolds
with $SU(3)$ structure and the corresponding prepotentials were derived in
\cite{MPS,eran}, but a detailed analysis in the vector multiplet sector of
these compactifications was not considered so far.

We begin with a short review of the compactification of M-theory on six
dimensional Calabi-Yau manifolds, and then proceed to the seven dimensional
case.

\subsection{M-theory compactifications on Calabi--Yau threefolds}\label{5dM}

In order to set the stage let us briefly recall the structure of the five
dimensional $N=2$ supergravity\footnote{By $N=2$ we mean the minimal amount of
  supersymmetry possible in five dimensions, which reduces to $N=2$ in four
  dimensions.} which arises from compactifying M-theory on Calabi--Yau
threefolds. Our discussion is based on
references~\cite{Gunaydin:1984ak,CCAF,AFT} but since we are only interested in
the vector multiplet sector we (largely) ignore the hypermultiplets in this
section.

The bosonic spectrum of eleven-dimensional supergravity is particularly simple
and consists only of the metric $\hat G_{MN}$ and a three-form potential $\hat
C_3$. (We use hats $\,\hat{}\,$ in order to denote the eleven-dimensional
quantities.) The eleven-dimensional action for these fields is given by
(setting the eleven dimensional Newton's constant to one)
\begin{equation}
  \label{S11}
  S_{11} = \frac12 \int \Big[ \hat R *1 - \frac12 \hat F_4 \wedge * \hat F_4
  - \frac16 \hat F_4 \wedge \hat F_4 \wedge \hat C_3 \Big] \; ,
\end{equation}
where $\hat F_4 = d \hat C_3$ is the field strength of the three-form potential
$\hat C_3$.

The five-dimensional vector fields arise from expanding
$\hat C_3$ in terms of harmonic $(1,1)$-forms on the Calabi-Yau.
More precisely we choose a basis $\omega_i$  of $H^{(1,1)}(CY_3)$ and expand
according to
\begin{equation}
  \label{C35dexp}
  \hat C_3 = A^i \wedge \omega_i + \ldots \; ,\qquad  i=1,\ldots , h^{(1,1)}\ ,
\end{equation}
where the $\ldots$ indicate further terms corresponding to scalar fields in
hypermultiplets.  One of the vector fields $A^i$ is identified with the
graviphoton while the other $(h^{(1,1)}-1)$ are members of vector multiplets.
Their (bosonic) superpartners correspond to K\"ahler deformations of the
Calabi--Yau metric. More precisely, one expands also the K\"ahler form $J$ in
terms of the basis $\omega_i$
\begin{equation}
  \label{klrmb}
  J\ =\ \vfive^i\, \omega_i\ ,
\end{equation}
such that the $\vfive^i$ parameterize the K\"ahler deformations.
In the five-dimensional low energy effective theory the $\vfive^i$
appear as scalar fields.
However, one of the K\"ahler moduli, the overall volume $\cK$,
is not part of any vector multiplet but instead is a member
of the universal hypermultiplet. The remaining $(h^{(1,1)}-1)$ moduli
are the scalar fields in vector multiplets.

Inserting \eqref{C35dexp} and \eqref{klrmb} into \eqref{S11} and integrating
over the Calabi-Yau manifold results in the five-dimensional $N=2$ effective
action (for the bosonic fields that are not in hypermultiplets)\footnote{Here
  we only give the final result and refer the reader for further details to
  \cite{CCAF,AFT}.}
\begin{equation}
  \label{5daction}
  S_5 = \int \left[ \tfrac12 R_5 *\mathbf{1}
  - g^{(5)}_{\alpha\beta}\, d \varphi^\alpha \wedge * d\varphi^\beta -
  \left. \tfrac14 g_{ij}\right |_{\cK=1} F^i \wedge * F^j - \tfrac1{12}
  \cK_{ijk} F^i \wedge F^j \wedge A^k \right ]\; ,
\end{equation}
where $F^i = dA^i$ and $\cK_{ijk}$
are intersection numbers of the Calabi-Yau defined by the integral
\begin{equation}
  \label{Khatdfb}
  {\cK}_{ijk} = \int_{CY_3} \omega_i \wedge \omega_j \wedge \omega_k
  \; .
\end{equation}

To explain the other couplings in this action we need to be more explicit
about the separation of the overall volume modulus $\cK$ from the other
K\"ahler moduli. Since the volume modulus is part of the universal
hypermultiplet, it should not mix with the other quantities describing the
vector multiplet moduli space. Therefore, all the terms in the vector
multiplet action \eqref{5daction} are evaluated on a hypersurface of constant
$\cK$ which we choose as $\cK=1$.  This is precisely the meaning of the matrix
of gauge couplings $\left . g_{ij} \right|_{\cK=1}$ in the action
\eqref{5daction}, which is equal to the metric on the K\"ahler moduli space
\cite{Strominger}
\begin{equation}
  \label{defgb}
  g_{ij} = \frac1{4 \cK} \int_{CY_3} \omega_i \wedge *\omega_j
    = - \frac{1}{4 \cK} \left( \cK_{ij} - \frac{\cK_i \cK_j}{4 \cK}
   \right) \; ,
\end{equation}
evaluated on the hypersurface $\cK=1$.\footnote{The same metric $g_{ij}$ will
  also appear in the four-dimensional effective action which we discuss in the
  next section. In this case it is the metric on a complex special K\"ahler
  manifold, since in $d=4$ the scalar fields in the vector multiplets are
  complex and furthermore they necessarily span a special K\"ahler manifold .}
Here the Calabi--Yau volume, $\cK$, is defined as
\begin{equation}
  \label{CYvolb}
  \cK = \tfrac16 \int_{CY_3} J \wedge J \wedge J = \tfrac16 \cK_{ijk} \vfive^i
  \vfive^j \vfive^k \; ,
\end{equation}
and we also abbreviated
\begin{equation}\begin{aligned}
  \label{cKdefsb}
  \cK_i  =&  \int_{CY_3} \omega_i\wedge J\wedge J =
  \cK_{ijk} \vfive^j \vfive^k\ ,\\
  \cK_{ij}  =&  \int_{CY_3} \omega_i\wedge\omega_j\wedge J =
  \cK_{ijk} \vfive^k \; .
\end{aligned}
\end{equation}

Finally let us discuss the kinetic terms of the scalar fields in the action
\eqref{5daction}. Let us denote by $\varphi^\alpha$ the $(h^{(1,1)}-1)$ scalar
fields which parameterize the hypersurface $\cK=1$. The metric
$g^{(5)}_{\alpha\beta}$ which appears in \eqref{5daction} is therefore the
induced metric on that hypersurface, which is given by
\cite{Gunaydin:1984ak,AFT}
\begin{equation}
  \label{5dmetric}
  g^{(5)}_{\alpha\beta}\ = \
  g_{ij}\ \frac{\partial \vfive^i}{\partial\varphi^\alpha}
  \frac{\partial \vfive^j}{\partial\varphi^\beta} \big|_{\cK=1}
  \; , \qquad \alpha,\beta = 1, \ldots, h^{(1,1)}-1\ .
\end{equation}

For the purpose of our paper it is of interest to also discuss possible
(global) isometries of the moduli space of the scalars in the vector
multiplets.  Following \cite{Gunaydin:1984ak} let us consider the
infinitesimal linear transformations
\begin{equation}
  \label{vgtrn}
  \vfive^i \to \vfive^i - \epsilon \M_j^i \vfive^j \ ,
\end{equation}
where the  $M^i_j$ are constant and elements of a Lie Algebra.
Since the space of scalar fields in vector multiplets
is defined on the hypersurface $\cK=1$
the transformation \eqref{vgtrn} is constrained by the requirement
\begin{equation}
  \label{delK}
  \delta \cK = 0 \; .
\end{equation}
Inserting \eqref{vgtrn} into \eqref{CYvolb} one arrives at
\cite{Gunaydin:1984ak}
\begin{equation}
  \label{conn}
  \M_i^l {{\cK}}_{jkl} + \M_j^l {{\cK}}_{kil} + \M_k^l
  {{\cK}}_{ijl} = 0 \ ,
\end{equation}
which states that $\cK_{ijk}$ is an invariant tensor of the Lie Algebra.
Inserting \eqref{vgtrn} into \eqref{cKdefsb} and \eqref{defgb}
one also computes
\begin{equation}
  \label{cKtr}
  \delta \cK_{ij}  =   \epsilon \M_i^k \cK_{kj} + \epsilon \M_j^k \cK_{ik} \;
  , \qquad
  \delta \cK_i  =   \epsilon \M_i^j \cK_j \; ,
\end{equation}
and
\begin{equation}
  \label{ggtr}
  \delta g_{ij} =  \epsilon \M_i^k g_{kj}  + \epsilon \M_j^k g_{ik} \; .
\end{equation}

By assigning the transformation law \eqref{vgtrn} also to the $A^i$ one
immediately sees the invariance of the last two terms in the action
\eqref{5daction}.  The invariance of the second term in \eqref{5daction} is
less obvious but has been established in \cite{Gunaydin:1984ak}. A quick
intuitive argument goes as follows. The full kinetic term on the Calabi--Yau
moduli space of K\"ahler deformations $g_{ij} \partial_\mu \vfive^i
\partial^\mu \vfive^j$ is clearly invariant under \eqref{vgtrn} and
\eqref{ggtr}. Since the second term in \eqref{5daction} differs from the one
above by a kinetic term for the volume modulus $\partial_\mu \cK \partial^\mu
\cK$, which is trivially invariant due to \eqref{delK}, it follows that the
the kinetic term for the K\"ahler moduli parameterizing the hypersurface
$\cK=1$ is also invariant under the transformation \eqref{vgtrn}.  Therefore
the action \eqref{5daction} has a global symmetry for any $M_j^i$ which solves
the constraint \eqref{conn}.

For generic $\cK_{ijk}$, eq.~\eqref{conn} has no solutions, or in other words,
a generic $\cK_{ijk}$ is not an invariant tensor of any Lie Algebra. Let us
therefore turn to a specific situation where global isometries do arise, which
will be used in the next subsections.  The case that we will discuss in detail
is the special class of $K3$-fibred Calabi-Yau threefolds (over a ${\bf P_1}$
base) \cite{KLM}. If we denote by $\vfive^1$ the volume of the base, then for
this class of manifolds the intersection numbers obey $\cK_{11i} = 0$.
Furthermore, if the ${\bf P_1}$ is taken large, i.e.\ $\vfive^1 \gg
\vfive^{i\neq1}$ for fixed $\cK$, then the moduli space is the scalar manifold
\cite{Gunaydin:1984ak,CCAF,AFT}
\begin{equation}
  \label{g5d}
  M_V = SO(1,1)\times \frac{SO(1,h^{(1,1)}-2)}{SO(h^{(1,1)}-2)}\ .
\end{equation}
The isometry group of this space is $SO(1,1)\times SO(1,h^{(1,1)}-2)$ and in
section \ref{K3fibred} we discuss in detail the corresponding solutions of
\eqref{conn}. A discrete subgroup of this isometry group,
$SO(1,h^{(1,1)}-2,{\bf Z})$, is known as the U-duality group which is an exact
symmetry of these compactifications.

\subsection{Seven dimensional manifolds with $SU(3)$ structure}
\label{SU3}
In the previous section we briefly reviewed Calabi-Yau compactifications of
M-theory. Let us now turn to compactifications on seven-dimensional manifolds
$X_7$ with $SU(3)$ structure. They can be characterized by a triplet of
globally defined and $SU(3)$-invariant tensors $\{V, J, \Omega\},$ where $V$
is a one-form, $J$ is a two-form and $\Omega$ is a three-form
\cite{waldram,DAP}.  This triplet is constrained to satisfy the compatibility
relations
\begin{equation}
\begin{aligned}
  \label{SU3con}
 & J \wedge J \wedge J  =  \tfrac{3 i}{4}\ \Omega \wedge \bar \Omega \
  , \\
  & \Omega \wedge J  =  V \lrcorner J\ =\ V \lrcorner \Omega\ =\ 0\ ,
\end{aligned}
\end{equation}
where $\lrcorner$ denotes contraction of indices.

Due to the existence of the one-form $V$, one can
define an almost product structure, in that locally the metric can be
split as
\begin{equation}
  \label{metricsplit}
  ds_7^2 (y,z) =  ds_6^2 (y,z) + V^2(y,z)\ ,
\end{equation}
where $y$ are the coordinates of the six-dimensional component $Y_6$ and $z$
is the coordinate of the one-dimensional component.  On $Y_6$ the two-form $J$
defines an almost complex structure (by raising one index with the metric) and
it is a $(1,1)$-tensor with respect to it. Similarly $\Omega$ is a $(3,0)$
form, and together they define the standard $SU(3)$ structure on a
six-dimensional space.

The manifold $X_7$ can be characterized by the non-vanishing intrinsic
torsion classes. They are defined by $dV, dJ,$ and $d\Omega$, and can
be decomposed into irreducible $SU(3)$ representations. One finds
13 torsion classes denoted $R$, $c_{1,2}$, $V_{1,2,3}$, $W_{1,2}$,
$A_{1,2}$, $T$ and $S_{1,2}$ in \cite{DAP}, defined by
\begin{eqnarray}
  \label{su3torsion}
  dV &=& R J + \bar{W_1} \lrcorner \Omega + W_1 \lrcorner \bar{\Omega} + A_1 +
  V \wedge V_1 \; , \nn \\
  dJ &=& \tfrac{2i}{3}\left( c_1 \Omega - \bar{c_1} \bar{\Omega} \right) + J
  \wedge V_2 + S_1 + V\wedge \left[ \tfrac{1}{3} \left( c_2 + {\bar c}_2\right)
  J + \bar{W_2}\lrcorner \Omega + W_2 \lrcorner \bar{\Omega} + A_2 \right]
  ,\ \nn\\
  d\Omega &=& c_1 J \wedge J + J \wedge T + \Omega \wedge V_3 + V \wedge
  \left[c_2 \Omega - 2 J \wedge W_2 + S_2 \right] \; .
\end{eqnarray}

As we already stated we do not compactify on generic $SU(3)$ structure
manifolds with all torsion classes non-zero. Instead we focus on manifolds
which can be viewed as Calabi--Yau threefolds $CY_3$ fibred over a circle
$S^1$.  With these specifications our setup is closely related to the case of
a six-dimensional torus $T^6$ fibred over a circle. Such backgrounds were
discussed in detail in~\cite{Hull,KM} and in the following we can draw on
their results.

We parameterize the $S^1$ direction by the coordinate $z \in [0, 1)$, while the
radius of the circle is given by the value of the dilaton $e^{\phi}$, where $V
= e^{\phi} dz$.  We further constrain the fibration such that when going
around the $S^1$ only the second cohomology $H^{(1,1)}(CY_3)$ is twisted by a
matrix $\gamma$, while the third cohomology $H^{3}(CY_3)$ is unaffected. In
this way we ensure that the hypermultiplet sector, which is governed by
$H^{3}(CY_3)$, coincides with that of a $CY_3\times S^1$ compactification.  On
the other hand, as we saw in the previous section, the vector multiplets are
determined by $H^{(1,1)}(CY_3)$ and hence they do feel the twisting.

As in the previous section we denote the elements of $H^{(1,1)}$ by $\omega_i$
but now they also depend on the circle coordinate $z$, or in other words we
have a set of $\omega_i(y,z)$. However, the structure of the fibration is not
arbitrary but constrained by a consistency condition.  If we choose a specific
basis at (say) $z=0$, it rotates as we move in the $z$ direction. After a full
circle, the $\omega_i$ must come back to an equivalent theory, i.e., the 5
dimensional theory returns to itself up to a discrete U-duality transformation
\cite{Hull}.\footnote{By U-duality we broadly refer to the group of discrete
  gauge transformations of the theory. We implicitly assume that all discrete
  global symmetries are actually gauged \cite{Dine:2004dk}.}
We already briefly discussed the U-duality group $\Gamma({\bf Z})$
of M-theory compactified on $CY_3$
at the end of the last subsection and here it appears as
the group of  monodromies
as we go around the circle\footnote{In
the last section we  noted that for
compactifications which have a heterotic dual
the U-duality group is  $\Gamma({\bf Z}) = SO(1,h^{(1,1)}-2,{\bf Z})$,
but the analysis of
this section holds for arbitrary $\Gamma({\bf Z})$.}
\begin{equation}
  \label{ofinite0}
  \omega_i \to \gamma^j_i\, \omega_j \ ,\qquad
  \gamma^i_j \in \Gamma({\bf Z})\ .
\end{equation}

In principle, only this global information exists. However, it is
convenient to choose an infinitesimal form of this relation by
twisting the basis $\omega_i$ by a constant matrix, $M^i_j$, in the
continuous group $\Gamma({\bf R})$ as we go along the circle.
In this case we write
\begin{equation}
  \label{gammainf}
  \gamma = e^{M},\qquad M \in \Gamma(\mathbf{R}) \; ,
\end{equation}
and the infinitesimal version of \eqref{ofinite0} becomes
\begin{equation}
  \label{infntsa}
  \omega_i(y,z+\epsilon) =\, \omega_i(y,z) + \epsilon \M_i^j \omega_j(y,z) \; .
\end{equation}
Since on the Calabi--Yau slice the $\omega_i$ continue to be
harmonic (and therefore closed), \eqref{infntsa} can also
be expressed by the differential relation
\begin{equation}
   \label{do}
   d \omega_i = \M_i^j\, \omega_j \wedge dz \; .
\end{equation}
Equations \eqref{infntsa} and \eqref{do} hold whenever the monodromy
is evenly distributed along the $S^1$.  This turns out to be useful
for carrying out a KK reduction even when there is no continuous
isometry.  In this case \eqref{infntsa} will in general not be a
solution of the equations of motion, but it is still a useful ansatz
for analyzing the compactification.  A specific example where this
ansatz gives a solution arises when we consider degenerations of the
Calabi-Yau compactification in which a continuous version of the
U-duality appears as an approximate global symmetry $\Gamma({\bf
  R})$. As we discussed at the end of section~\ref{5dM} this situation
occurs, for example, when the base in the $K3$-fibred $CY_3$ is
large. In this case the matrix $M$ satisfies \eqref{conn}, and
\eqref{infntsa} expresses a translation invariance along the $S^1$.

However, generically the full theory does not have the continuous
symmetry and the only real information is the global data in the
monodromy $\gamma$. The approach that we will take is first to discuss
the situation with a continuous symmetry, obtain a quantitative
understanding of what this process of twisting does, and afterwards
indicate (in section \ref{break}) why going away from this limit does
not change the qualitative picture.

Let us define the Calabi-Yau intersection numbers exactly as in \eqref{Khatdfb},
but now with $z$-dependent $\omega_i(y,z)$.
In this case the $\cK_{ijk}$  can a priori also be $z$-dependent. However,
inserting \eqref{infntsa} into \eqref{Khatdfb}, we see
that precisely when there
is an isometry the $z$-dependence cancels out due to \eqref{conn}.
Note that the fact that the same matrix $M_i^j$
appears in \eqref{conn} and \eqref{infntsa} establishes the
connection between an isometry in the space-time effective theory
and the translational symmetry of the fibration of the $(1,1)$-forms
along the $S^1$-circle.

In the Kaluza-Klein reduction which we perform in the next section we
encounter the seven-dimensional integral
\begin{equation}
  \label{intno}
  \hat\cK_{ijk} = \int_{X_7} \omega_i \wedge \omega_j \wedge \omega_k \wedge dz
  \; ,
\end{equation}
which are the intersection numbers defined on the entire $X_7$. They coincide
with the ${\cK}_{ijk}$ 
precisely when \eqref{conn} holds. In this case the ${\cK}_{ijk}$ are
$z$-independent, and thus the integral in \eqref{intno} trivially
factorizes. Note
that the condition \eqref{conn} also arises in this case
from the requirement of global consistency of \eqref{do}
\begin{equation}
  \label{conseq}
  \int_{X_7} d( \omega_i \wedge \omega_j \wedge \omega_k) = 0\; .
\end{equation}

It is also useful to note that all the other Calabi--Yau moduli space
quantities defined in equations \eqref{defgb}, \eqref{CYvolb} and
\eqref{cKdefsb} have, due to \eqref{conn}, similar definitions in terms
of seven-dimensional integrals. In particular the Calabi--Yau volume
can also be defined as
\begin{equation}
  \label{volCY7}
  \cK = \tfrac16 \int_{X_7} J \wedge J \wedge J \wedge dz \; .
\end{equation}
The volume of the full seven-dimensional manifold $X_7$ differs
from this one by a dilaton factor, which, when the dilaton is
independent of the $X_7$ coordinates is equal to
\begin{equation}
  \label{volX}
  \hat \cK = \tfrac16 \int_{X_7} J \wedge J \wedge J \wedge V = e^{\phi} \cK \; .
\end{equation}

In analogy with \eqref{klrmb} we  expand $J$ according to
\begin{equation}
  \label{Jdef}
  J\ =\ v^i\, \omega_i(y,z)\ ,
\end{equation}
where the $v^i$ are again constant but now there is a $z$-dependence in
$\omega_i$.
The $v^i$  will appear as scalar fields in the
four-dimensional effective action.
Note that
$J$ is not invariant under translation in the
$z$-direction, but it comes back to itself when we go all the way
around the circle. This follows from the fact that
we identify the manifold under
$z\rightarrow z+1$ together with \eqref{ofinite0}. As a consequence $J$ is
globally defined on $X_7$.

As we will see in the next subsection, it is the $z$-dependence of the
$\omega_i$ in
\eqref{Jdef} which generates mass terms for the
fields $v^i$ in the four-dimensional effective action. Let us note that
this can also be seen from a Scherk-Schwarz point of view \cite{SS}
where one first
compactifies to five dimensions on Calabi--Yau manifolds as in the
previous section and then, in the subsequent compactification to four dimensions,
gives the five-dimensional scalar fields $\vfive^i$ a monodromy as one moves around the circle
such that their $z$-dependence is given by 
$\vfive^i(z + \epsilon) = \vfive^i(z) + \epsilon M_j^i\,
\vfive^j(z)$. Thus the
relation between $\vfive^i$ and $v^i$ is simply $\vfive^i(z) = (e^{z
  M})_j^i\, v^j$.

Inserting eq.~\eqref{Jdef} into eq.~\eqref{volCY7} and 
using eq.~\eqref{intno} and $\hat\cK_{ijk} = \cK_{ijk}$ we obtain
$\cK=\tfrac16 \cK_{ijk}v^iv^jv^k$ exactly as in \eqref{CYvolb}, but now
in terms of the parameters $v^i$ instead of $\vfive^i$.
Similarly, the metric on the moduli space of K\"ahler deformations can
be defined as
\begin{equation}
  \label{defg}
  g_{ij} = \frac1{4 \cK} \int_{X_7} \omega_i \wedge *\omega_j \; ,
\end{equation}
with no dilaton prefactor, which is in agreement with the metric Ansatz
\eqref{11dg} we shall consider in the next section. One can show that
it coincides with the metric given in eq.~\eqref{defgb} with the replacement
$\vfive^i\to v^i$ in \eqref{CYvolb} and \eqref{cKdefsb}.

Before we turn to the details of the KK-reduction let us determine the
non-trivial torsion classes in \eqref{su3torsion} for the fibration
characterized by eq.~\eqref{infntsa} or equivalently by eq.~\eqref{do}. Using
the expansion \eqref{Jdef} with the forms $\omega_i$ satisfying
\eqref{do} we find
\begin{equation}
  \label{dJ}
  dJ = v^i M_i^j\, \omega_j\wedge dz\ ,
\end{equation}
which shows that the $\M_i^j$ parameterize the non-vanishing intrinsic
torsion. Comparison with \eqref{su3torsion} reveals that the only
torsion classes which can be non-trivial are $A_2$ and $\mathrm{Re}~c_2$. 
Actually, for the case at hand, it can be shown that
$\mathrm{Re}~c_2$ vanishes and the only torsion class which is
present is $A_2$. This can be seen by writing \eqref{dJ} in components and
contracting with $J^{mn}$. Using the $SU(3)$ structure consistency
relation $V_m J^{mn} = 0$, \eqref{SU3con}, and the fact that $A_2$ in
\eqref{dJ} is primitive, i.e. $(A_2)_{mn} J^{mn} = 0$, leaves us with the
following relation for $\mathrm{Re}~c_2$
\begin{equation}
  \label{Rec}
  \mathrm{Re}~c_2 \sim v^i \M_i^j (\omega_j)_{mn} J^{mn} \; .
\end{equation}
For Calabi--Yau manifolds, the contraction of the $(1,1)$ forms
$\omega_j$ with $J$ was computed in \cite{Strominger} and shown to be
proportional to $\cK_{jkl} v^k v^l$. Inserting this into the above
equation, the vanishing of the torsion class $\mathrm{Re}~c_2$ is
simply a consequence of the constraint \eqref{conn}.
Note that for $\M=0$ the two-form $J$ is
closed, the fibration is trivial and $X_7$ is the product manifold
$CY_3\times S^1$.

\subsection{Kaluza-Klein reduction of M-theory  on $X_7$}
\label{sec:KK}

We can now proceed with one of the main parts of this paper, namely the
compactification of M-theory, or rather eleven-dimensional supergravity, on
seven-dimensional manifolds with $SU(3)$ structure. As explained before, we
concentrate on the vector multiplet sector and ignore the hypermultiplets in
our analysis.

The starting point is the eleven-dimensional action \eqref{S11}.
Since on seven-dimensional manifolds with $SU(3)$ structure we can define an
almost product structure we consider the following Ansatz for the metric
\begin{equation}
  \label{11dg}
  G_{MN} = \left(
    \begin{array}[h]{ccc}
      e^{4\phi/3} \left(\frac{1}{\cK} G_{\mu \nu} +  A^0_\mu A^0_\nu \right) &
      0 & - e^{4\phi /3} A^0_\mu \\
      0 & e^{-2\phi/3} G_{mn} & 0 \\
     -e^{4\phi /3} A^0_\nu & 0 & e^{4 \phi /3}
    \end{array}
    \right) \; ,
\end{equation}
where $ G_{\mu \nu}$ denotes the 4d metric, $G_{mn}$ is the metric on the
Calabi--Yau manifold,
$A^0_\mu$ is the $4d$ graviphoton and $\phi$ the dilaton.\footnote{The above
Ansatz includes only  zero modes, and therefore we omitted the
off-diagonal components which involve one-forms on $CY_3$, since they
lead to massive excitations.}
The scalar fields arising from the Calabi-Yau metric correspond to the
deformations of $J$ which we denoted by $v^i$ in \eqref{Jdef}, as well
as the deformations of $\Omega$. The dilaton factors are chosen in such a way
that we end up in the four dimensional Einstein frame. The factor $1/\cK$ --
with $\cK$ defined in \eqref{volCY7} -- in
front of the
four-dimensional metric has been introduced to account for the additional
Calabi--Yau volume factor which appears in front of the Einstein-Hilbert term
after performing the integral over the internal manifold.

Next we expand the three-form potential according to
\begin{equation}
  \label{expC}
  \hat C_3 = \tilde C_3 + B \wedge dz + \tilde A^i \wedge \omega_i + b^i
  \omega_i \wedge dz +
\ldots
 \; ,
\end{equation}
where $\tilde C_3$ is a three-form in four dimensions, $B$ is a
two-form, $\tilde A^i$ are vector fields and $b^i$ are scalars. The $\ldots$
stand for additional scalar fields that arise when  $\hat C_3$ is expanded in
a basis of $H^3(CY_3)$, which, together with the complex structure
deformations, the dual of $B$, and the dilaton $\phi$, fill out
$h^{(1,2)} +1$ hypermultiplets, and we omit them from our further
discussion.\footnote{The
  couplings of the hypermultiplets in the $N=2$ low energy effective
  action can be found, for example, in \cite{FS,LM2}.}
We do keep the gravity
multiplet which includes the graviton and the graviphoton $A^0$, and the
$h^{(1,1)}$
vector multiplets which include the vector fields $\tilde A^i$ and the
complex scalars $x^i = b^i + iv^i$. Note that compared to the five-dimensional
case discussed in section \ref{5dM},  there is an additional
vector multiplet and the K\"ahler moduli are complexified.
Thus, in the four-dimensional effective action all K\"ahler moduli,
including the Calabi--Yau volume, are in vector multiplets.

In the compactification process it is useful to keep track of the
isometries of the internal manifold $X_7$ since they become gauge
transformations in the effective theory. Let us first recall the
situation for compactifications on $CY_3 \times S^1$.  In this case
there is an isometry corresponding to constant shifts $z\to
z+\epsilon$ of the $S^1$ coordinate.
Promoting the parameter to be space-time dependent
$\epsilon\to\epsilon(x^\mu)$, the compactification Ansatz given in
eqs.~\eqref{11dg} and \eqref{expC} changes. Keeping $\hat C_3$ and the
ten-dimensional line-element $ds^2_{10}$ invariant induces the local
gauge transformations
\begin{equation}
  \label{11diff}
  A^0 \to  A^0 + d \epsilon \; , \qquad
  \tilde C_3 \to \tilde C_3 - B \wedge d \epsilon \; , \qquad
  \tilde A^i \to \tilde A^{i} - b^i d \epsilon \; .
\end{equation}
However, the fact that the fields $\tilde C_3$ and $\tilde A^i$ transform is an
artefact of the expansion \eqref{expC} and one can define the
gauge-invariant fields
\begin{eqnarray}
  \label{defCA}
  C_3 = \tilde C_3 + B \wedge A^0 \; , \qquad A^i = \tilde A^i + b^i A^0 \; .
\end{eqnarray}

In the case of a non-trivial fibration of the Calabi--Yau over the
circle, as considered in eq.~\eqref{infntsa}, these fields are no longer
invariant. However, the main 
property of eq.~\eqref{defCA} is that the transformations of $C_3$ and
$A^i$ do not contain the derivative of the transformation parameter
$\epsilon$, and therefore we will keep the same definitions in the
following. Using eqs.~\eqref{infntsa} and \eqref{expC} we can easily see
that the fields $A^i$ and $b^i$ arising from the expansion of $\hat
C_3$ acquire a non-trivial gauge transformation, but the
transformation law of the graviphoton is unchanged.

Exactly as for $\hat C_3$, we also need to keep $J$, defined in
eq.~\eqref{Jdef}, 
gauge-invariant. Since the basis of $(1,1)$ forms $\omega_i$ changes according
to eq.~\eqref{infntsa}, we need to assign a transformation law similar to
\eqref{vgtrn} also to the fields $v^i$. Another way of saying this is that our
background is not invariant under arbitrary shifts of $z$, but, as shifts of
$z$ are gauge symmetries, we must assign a transformation law to the $v^i$.
Altogether we thus have
\begin{equation}
  \begin{aligned}
    \label{epsgtr}
    A^0 & \to  A^0 + d \epsilon \; ,\qquad
    A^i  \to  A^i - \epsilon \M_j^i A^j \; , \\
    v^i & \to  v^i - \epsilon \M_j^i v^j\; , \qquad   b^i  \to  b^i -
    \epsilon \M_j^i b^j \; .
  \end{aligned} 
\end{equation} 
Note that unlike the $N=2$ gauged supergravities encountered so far in
string compactifications, the symmetry \eqref{epsgtr} is not necessarily
a Peccei-Quinn shift symmetry which is usually gauged in these cases.
Moreover, this gauge symmetry is generically spontaneously broken due to
the non-vanishing vacuum expectation values of the K\"ahler moduli $v^i$.

In addition, the four-dimensional effective theory sees the remnant of
the three-form gauge invariance  $\hat C_3 \to \hat C_3 + d
\Lambda_2$ which is manifest in the action \eqref{S11}. Choosing $\Lambda_2 = \eta^i \omega_i$ and using
\eqref{do} we obtain the following transformation laws %
\begin{equation}\begin{aligned}
  \label{etagtr}
 A^0 & \to  A^0 \; ,\qquad
  A^i  \to  A^i + d \eta ^i + \M_j^i \eta^j A^0 \; , \\
  v^i & \to  v^i \; , \qquad b^i  \to  b^i + \M_j^i \eta^j \; .
\end{aligned}\end{equation}
The parameters $(\epsilon, \eta^i)$ together form $h^{(1,1)}+1$ local gauge
parameters. From the transformations displayed in \eqref{epsgtr} and
\eqref{etagtr} we already see the non-Abelian character of the gauge
transformations when $M \neq 0$.

To derive the four-dimensional action we insert eqs.~\eqref{11dg} and
\eqref{expC} into \eqref{S11} and perform the integrals over the internal
manifold. Let us first concentrate on the last two terms in the action
\eqref{S11}, and postpone the compactification of the Ricci scalar to the end
of this section.

To make our task easier let us first compute the field strength $\hat F_4$ by
taking the exterior derivative of eq.~\eqref{expC}. Using \eqref{do} and the
definitions \eqref{defCA} we find
\begin{equation}\begin{aligned}
  \label{F4}
  \hat F_4  = d \hat C_3 = \ & (d C_3 - B \wedge F^0) + H \wedge (dz - A^0) \\
  &  + (F^i - b^i F^0) \wedge \omega_i + D b^i \wedge \omega_i \wedge (dz -
  A^0) + \ldots \, ,
\end{aligned}\end{equation}
where we defined
\begin{equation}
  \label{cd}
    F^0 =  d A^0 \ , \qquad
    F^i  =  d A^i - \M_j^i\, A^j \wedge A^0 \; , \qquad
    D b^i  =  d b^i - \M_j^i\, ( A^j - b^j A^0) \; .
\end{equation}
The reason we have formally performed the expansion in the forms $dz -
A^0$ is that in this basis the metric \eqref{11dg} is block diagonal,
and therefore in computing $(\hat F_4)^2$ only the square of the
individual terms in \eqref{F4} will appear and no mixed terms will be
present. Note that in four dimensions $C_3$ is not a dynamical field and
therefore we will discard its contribution in the following. In general, a
proper dualization should be performed, but this has implications only on the
hypermultiplet sector and is therefore not of interest for us. 
With these things in mind we obtain
\begin{equation}
  \label{kin4}
  \int_{X_7} \hat F_4 \wedge * \hat F_4 =
  e^{-4 \varphi} H_3 \wedge * H_3 + 4 \cK g_{ij} (F^i - b^i
  F^0) \wedge * (F^j - b^j F^0) + 4 g_{ij} Db^i \wedge * Db^j + \ldots \; ,
\end{equation}
where the metric $g_{ij}$ was defined in eq.~\eqref{defg} and $\varphi$ denotes
the four dimensional dilaton defined as $e^{-2\varphi} = e^{- 2 \phi} \cK$.
For the Chern-Simons term in \eqref{S11} one finds after a straightforward but
somewhat lengthy calculation
\begin{equation}\begin{aligned}
  \label{FwF}
  \int_{X_7} \hat C_3 \wedge \hat F_4 \wedge \hat F_4\ =\ & 3 F^i \wedge F^j
  b^k \cK_{ijk} - 3 F^i \wedge F^0 b^j b^k \cK_{ijk} \\
& + F^0 \wedge F^0 b^i b^j
  b^k \cK_{ijk}
 + 2 \M_i^k A^i \wedge A^l \wedge F^j \cK_{jkl} \; ,
\end{aligned}\end{equation}
where $\cK_{ijk}$ are the Calabi--Yau intersection numbers defined in
\eqref{Khatdfb} which can also be obtained from \eqref{intno}.

Let us check explicitly that the individual terms in eq.~\eqref{kin4} are
invariant under the gauge transformations \eqref{epsgtr} and \eqref{etagtr}.
Under \eqref{epsgtr} the quantities
defined in eq.~\eqref{cd} transform as
\begin{eqnarray}
  \label{trDbF}
  \delta Db^i  =  -\epsilon \M_j^i Db^j  \; ,\qquad
  \delta F^i  =  -\epsilon \M_j^i F^j  \;,\qquad
  \delta F^0  = 0\ .
\end{eqnarray}
Together with the transformation \eqref{ggtr} of the moduli space metric, this
shows that the terms in 
eq.~\eqref{kin4} are (individually) invariant. Under
the gauge transformation \eqref{etagtr}, the covariant derivatives
$Db^i$ are invariant as can be checked from \eqref{cd}. The  field strengths
$F^i, F^0$ are not individually invariant, but the combinations
\begin{equation}
  \label{Finv}
  \check F^i = F^i - b^i F^0 \; ,
\end{equation}
which appear in \eqref{kin4}, are invariant. This completes the proof
of the gauge invariance of the expression \eqref{kin4}.

We can similarly check the gauge invariance of \eqref{FwF}. For the
transformation \eqref{epsgtr} it follows straightforwardly from
eq.~\eqref{trDbF} 
and the constraint \eqref{conn} that each term in \eqref{FwF} is invariant
individually. To check the invariance under the transformation \eqref{etagtr}
is also straightforward but a bit more tedious.
The important difference to note is
that the gauge invariance \eqref{etagtr} only holds for the sum of all the
terms in eq.~\eqref{FwF} but not for the individual terms. We come back to
this issue in section \ref{sec:N2}.

The next step is to compactify the Ricci scalar in the action
\eqref{S11}.
For $CY_3\times S^1$ the answer is well known \cite{CCAF} and yields
the kinetic terms for the moduli $v^i$, a contribution to the kinetic
terms of the graviphoton $A^0$ and the kinetic term for the dilaton.
For the case of a non-trivial fibration the moduli are charged under
the isometry of the circle and the corresponding
gauge transformation is given in \eqref{epsgtr}. This in turn leads to a
coupling of the moduli to the graviphoton and a scalar potential.
The generic formulae for this case are worked out in \cite{SS} and we
can borrow some of their results. One finds
\begin{equation}
  \label{Raction}
  \tfrac12 \int_{X_7} \hat R *1 = \tfrac12 R_4 * \mathbf{1} - g_{ij} Dv^i
  \wedge *Dv^j - \cK F^0\wedge * F^0 - d\varphi\wedge *d\varphi -V\ .
\end{equation}
This is a straightforward generalization of the result obtained in $CY_3
\times S^1$ compactifications, in that the
derivatives for the charged moduli are replaced by covariant derivatives
\begin{equation}
  \label{Dv}
  Dv^i = dv^i  + v^j M_j^i A^0 \ .
\end{equation}

The derivation of the scalar potential $V$ is less obvious and an explicit
calculation of the internal Ricci scalar is necessary. Note that this gives in
fact the only contribution to the potential as eqs.~\eqref{kin4} and
\eqref{FwF} 
contain no terms without four-dimensional derivatives. Let us therefore
compute the scalar curvature for the internal part of the metric which can be
read off from \eqref{11dg}
\begin{equation}
  \label{gint}
  G_{int} = e^{-2\phi/3} \left(
    \begin{array}[h!]{cc}
      G_{mn}(y,z)  & 0  \\
      0 & e^{2\phi} \\
    \end{array}
    \right) \; .
\end{equation}
{}From the seven-dimensional point of view, the overall dilaton factor is
irrelevant as this is just a constant, but it will be important for the
normalization of the potential. Using the fact that the Ricci tensor of the
Calabi--Yau slices vanishes we find
\begin{equation}
  \label{R7}
  R_7 = - e^{-4\phi /3} \Big[ \partial_z(G^{mn} \partial_z G_{mn} ) + \tfrac14
  (G^{mn} \partial_z G_{mn})^2 + \tfrac14 G^{mn} G^{pq} \partial_z G_{mp}
  \partial_z G_{nq} \Big] \; .
\end{equation}
In order to proceed we split the metric into a background piece
$G_{mn}^0$, which is
constant in $z$, and the moduli dependent part $\Delta G_{mn}$
which does depend on $z$ :
\begin{equation}
  \label{pert}
  G_{mn} = G_{mn}^0 + \Delta G_{mn} \; .
\end{equation}
As explained before, the fibration structure we consider is such that the
complex structure deformation sector is not influenced by the additional $z$
direction, and we are only interested in the dependence on the
K\"ahler moduli $v^i$. In complex coordinates they arise from
the $(1,1)$ components of the metric via
\begin{equation}
  \Delta G_{a \bar b} = - i v^i (\omega_i)_{a \bar b} \; , \qquad
  a,\bar b=1,2,3\ .
\end{equation}
Using eq.~\eqref{infntsa} we immediately find
\begin{equation}
  \label{ddg}
  \partial_z \Delta G_{a \bar b} = -i v^i \M_i^j (\omega_j)_{a \bar b} \; .
\end{equation}
{}From the fact that $\omega_j$ is a harmonic $(1,1)$-form on the
Calabi-Yau threefold, one shows, following Ref.~\cite{Strominger}, that
$G^{a \bar  b} (\omega_j)_{a \bar b} = \tfrac{i}2 \cK_j/\cK$, where
eqs.~\eqref{CYvolb} and \eqref{cKdefsb} were used. Combining this with
eq.~\eqref{ddg} gives
\begin{equation}
  G^{mn} \partial_z G_{mn} = \cK_{jkl} \M_i^j v^i v^k v^l =0 \; ,
\end{equation}
as a consequence of the constraint \eqref{conn}. Therefore the only
contribution to the four-dimensional potential comes from the last term in
\eqref{R7}. Inserting eq.~\eqref{ddg} into eq.~\eqref{R7} we arrive at
\begin{equation}
  \label{intR}
  \tfrac12 \int_{X_7} R_7 = - \tfrac14 e^{-4\phi/3} \M_i^k \M_j^l v^i v^j
\int_{X_7}\omega_k  \wedge * \omega_l \; .
\end{equation}
Using eqs.~\eqref{defg} and \eqref{11dg},
and taking into account the rescaling of the four-dimensional metric,
we finally obtain the potential (in the Einstein frame)
\begin{equation}
  \label{4dpot}
  V = \frac1\cK\, v^i v^j \M_i^k \M_j^l g_{kl} \; .
\end{equation}
%

\subsection{Consistency with $N=2$ supergravity}
\label{sec:N2}

In order to check the consistency with $N=2$ supergravity
(reviewed in appendix~\ref{sg}) we have to
write the resulting four-dimensional action  in the general form \eqref{agsg}.
Putting together eqs.~\eqref{kin4}, \eqref{FwF} and \eqref{Raction} we obtain
the action in four dimensions for the bosonic fields in the gravity and
vector- multiplets 
\begin{eqnarray}
  \label{S4}
  S  & = & \int_{M_4} \Big[\tfrac 12\, R *1 - g_{ij} Dx^i \wedge
  *D{\bar x}^{j} -V
  \\
  & & \qquad + \tfrac14 \,\mathrm{Im} \cN_{IJ} F^I \wedge *
  F^J + \tfrac14\,  \mathrm{Re} \cN_{IJ} F^I \wedge F^J - \tfrac16\, \M_i^l \cK_{jkl}
  A^i \wedge A^j \wedge dA^k\Big] \; , \nn
\end{eqnarray}
where the metric $g_{ij}$ was defined in \eqref{defg}. It
is a special K\"ahler metric derived from the K\"ahler
potential given in \eqref{Kspecial} for the prepotential
\begin{equation}
  \label{prep}
  \cF(X) = - \frac16\, \frac{\cK_{ijk} X^i X^j X^k}{X^0} \; .
\end{equation}
The $X^I, \ I=0, \ldots, h^{(1,1)}$, are  projective coordinates which are
related to the scalar fields via so-called special coordinates $x^i$ given
by
\begin{equation}
  \label{prcoord}
  x^i = \frac{X^i}{X^0} =  b^i + i v^i \; , \qquad i=1, \ldots h^{(1,1)} \; .
\end{equation}
The prepotential \eqref{prep} also determines the gauge coupling matrix $\cN$
via \eqref{Ndef}, and one finds
\begin{equation}\begin{aligned}
  \label{Nexp}
\mathrm{Re}\, {\cal  N}_{00} &= - \tfrac13\cK_{ijk} b^i b^j b^k\ , \qquad
\mathrm{Re}\, {\cal  N}_{i0} =  \tfrac12\cK_{ijk}  b^j b^k\ , \qquad
\mathrm{Re}\, {\cal  N}_{ij} = - \cK_{ijk} b^k\ , \\
\mathrm{Im}\, {\cal  N}_{00} &= - \cK(1 + 4 g_{ij} b^i b^j) \ , \qquad
\mathrm{Im}\, {\cal  N}_{i0} = 4\cK g_{ij}  b^j \ , \qquad
\mathrm{Im}\, {\cal  N}_{ij} =  -4 \cK g_{ij} \ .
\end{aligned}
\end{equation}

The field strengths in eq.~\eqref{S4} are given by
\begin{equation}
  \label{strcon}
  F^I = d A^I + \tfrac12 f^I_{JK}A^J\wedge A^K \ , \qquad\mathrm{with}\qquad
  f_{IJ}^0 = 0 = f_{ij}^k  \; , \qquad f_{i0}^j = - \M_i^j  \; ,
\end{equation}
while the covariant derivatives read
\begin{eqnarray}
  \label{kvA}
  Dx^i = dx^i - k^i_I A^I \ , \qquad\mathrm{with}\qquad  k_0^j = - x^k \M_k^j
  \; , \qquad k_i^j = \M_i^j \; .
\end{eqnarray}
These holomorphic Killing vectors can be obtained via \eqref{Pdef} from the
Killing prepotentials
\begin{equation}
  P_0 = - x^i M_i^j K_j \,, \qquad P_i = M_i^j K_j\ ,
\end{equation}
where
$K_j= \partial_j K$ is the first derivative of the K\"ahler potential.
The consistency of the non-Abelian gauge
algebra can be checked in that eq.~\eqref{commutator} is fulfilled and we have
\begin{equation}
  \label{algk}
  [k_i,k_j] = 0 = [k_0,k_0] \ , \qquad [k_i,k_0] = - M_i^j k_j \ ,
\end{equation}
corresponding to a semi-direct sum of two Abelian sub-algebras.
\footnote{We thank the referee of this paper for pointing
  out that \eqref{algk} is also a solvable Lie algebra (for a
  definition see, for example, \cite{Turin}).} Finally,
using \eqref{kvA} it is easy to see that the potential \eqref{4dpot} is
consistent with \eqref{pot}.

Except for the last term in eq.~\eqref{S4} everything looks like a standard
$N=2$ gauged supergravity as spelled out in ref.~\cite{N=2review}. The last
term is also known, and has to be introduced in the action (in order to make
it gauge-invariant) whenever the prepotential is not
invariant under the gauge transformations, but transforms into a second order
polynomial in $X$ with real coefficients \cite{dWvP}. Inserting the
transformation \eqref{etagtr} into the definition of the projective
coordinates \eqref{prcoord} we find that the prepotential \eqref{prep} changes
as
\begin{equation}
  \label{Ftr}
  \delta_{\eta} \cF = - \frac12 \eta ^i \M^l_i \cK_{ljk} X^j X^k   \; ,
\end{equation}
which is precisely of the form \eqref{dF} with
$C_{ijk} =\tfrac12\M^l_i \cK_{ljk}$. Note that for the specific structure
constants given in eq.~\eqref{strcon}, the last term of eq.~\eqref{dL}
vanishes, which explains why such a term is absent from eq.~\eqref{S4}.

Before we continue it is worthwhile to
stress that the vector multiplet geometry on the M-theory side
specified by the prepotential \eqref{prep} is exact,
since the ``dilaton'' (the radius of the M-theory circle) is part of a
hypermultiplet, and therefore cannot
correct this geometry. The same holds for the gauging as specified in
\eqref{kvA}.

\subsection{$K3$-fibred Calabi-Yau threefolds}
\label{K3fibred}

So far our discussion was generic, in that we did not specify the
intersection numbers $\cK_{ijk}$ and the matrix $M_i^j$. We did however
assume that the seven-dimensional $X_7$ is a fibred product of a
Calabi-Yau threefold $CY_3$ over a circle, and that the $CY_3$ is such
that a continuous isometry of the form \eqref{infntsa} exists.
In this section we discuss more concretely the specific case of
$K3$-fibred Calabi-Yau threefolds; type IIA string theory compactified
 on such threefolds is dual to heterotic string theory
compactified on $K3\times T^2$.

$K3$-fibred Calabi-Yau threefolds consist of $K3$ fibres
 over a ${\bf P_1}$ base \cite{KLM}.
The volume of the base in string units is identified with the
dilaton on the heterotic side. Furthermore, two additional
two-cycles in the $K3$, related to the heterotic torus, can be
singled out. Let us denote these three special cycles by $1$, $2$
and $3$, while the rest of the two-cycles are denoted by an index
$a$. In the limit of a large ${\bf P_1}$ base (i.e.\ large heterotic
dilaton) the prepotential
 \eqref{prep} becomes
\begin{equation}
  \label{IIAprep}
  \cF = \frac{X^1(X^2 X^3 - X^a X^a)}{X^0}
\end{equation}
and so the only non-vanishing intersection numbers for the Calabi--Yau
threefold are \cite{KLM}
\begin{equation}
  \label{intcy}
  \cK_{123} = -1 \; , \qquad \cK_{1ab} = 2 \delta_{ab} \; , \qquad
a,b=4,\ldots,h^{(1,1)}\ .
\end{equation}
Inserting eq.~\eqref{IIAprep} into eq.~\eqref{Kspecial} and computing the
corresponding K\"ahler metric one sees that this factorizes and
becomes the metric on the space
\begin{equation}
  \label{gIIA}
  M_V = \frac{SU(1,1)}{U(1)}\times \frac{SO(2,h^{(1,1)}-1)}{SO(2)\times
  SO(h^{(1,1)}-1)}\ . 
\end{equation}
The first factor is spanned by the coordinate $x^1$ which parameterizes
the volume of the ${\bf P_1}$ base, while $x^2, x^3$ and $x^a$ span
the second factor. We immediately see that $M_V$ has the
continuous isometry group $SU(1,1)\times SO(2,h^{(1,1)}-1)$.
As discussed above, in the same limit the five dimensional vector multiplet moduli
space has the continuous isometry group $SO(1,1)\times SO(1,h^{(1,1)}-2)$.

As a consequence, we expect that the constraint \eqref{conn} has
non-trivial solutions. Indeed, solving eq.~\eqref{conn} for the torsion
parameters $\M_i^j$, given the intersection numbers \eqref{intcy}, we find
that one can choose to express all matrix elements in terms of $\tfrac12
(h^{(1,1)} - 1) (h^{(1,1)}-2) +1$ independent parameters
\begin{equation}
  \label{indpar}
  m_2   \equiv \M_2^2  \; , \qquad m_a \equiv  \M_a^2 \; , \qquad
  m_3  \equiv \M_3^3 \; , \qquad \tilde m_a \equiv \M_a^3 \; , \qquad
  m_b^a \equiv -  \M_a^b  \  ,
\end{equation}
where $m^a_{b} = -m^b_{a}$.
The other matrix elements are then given by
\begin{equation}
  \label{deppar}
  \begin{aligned}
    \M_2^a  =  \tfrac12 \tilde m_a  \; , & \qquad \M_3^a = \tfrac12 m_a \; ,
    \qquad  \M_a^a = - \tfrac12 M_1^1\ =\ \tfrac12 (m_2 +  m_3) \;
    ,\\
    & M_1^{2,3} = M_1^a =  M_a^{1} = M_{2,3}^1  = M_2^3= M_3^2 = 0
\ .
  \end{aligned}
\end{equation}
Note that these solutions describe the mixing of $SO(1,1)\times
SO(1,h^{(1,1)}-2)$ into the gauge symmetry, i.e, we have accounted
for the most general monodromy allowed on the circle. However, this
is not the most general global symmetry of the four dimensional theory, which
can be as large as $SU(1,1)\times SO(2,h^{(1,1)}-1)$. In
section~\ref{hetsect} we discuss how the parameters in \eqref{deppar}
are related to the dual heterotic background.
Before we do so let us return to the situation where the
${\bf P_1}$-base is not necessarily large.

\subsection{Breaking the continuous isometry }\label{break}

So far our analysis assumed that the Calabi-Yau moduli space has a
continuous isometry, or in other words that $\cK_{ijk}$ are such
that eq.~\eqref{conn} has a solution. As we saw this is indeed the case
for $K3$-fibred Calabi-Yau manifolds in the large
${\bf P_1}$ limit where the moduli space has
a continuous $SO(1,h^{(1,1)}-2)$ symmetry. However, this symmetry
is broken (for example, by
non-zero intersection numbers $\cK_{abc}$ or $\cK_{23a}$)
 to a
discrete subgroup $\Gamma({\bf Z}) = SO(1,h^{(1,1)}-2,{\bf Z})$ (which
is the T-duality group of the heterotic string) for finite
${\bf P_1}$ volume. As
we discussed before, the only information that we can really specify
at finite ${\bf P_1}$ volume is an element $\gamma_i^j \in \Gamma({\bf Z})$,
which rotates the $(1,1)$-forms $\omega_i$ as described in
section~\ref{SU3}. Of course the absence of the continuous isometry
also holds for compactifications on $CY_3\times S^1$ without any
monodromy; in this case the corresponding continuous symmetry
is broken to a discrete subgroup $\Gamma'(\bf Z)$, which is the
T-duality group of the dual heterotic string on $K3\times
T^2$. Furthermore in
four dimensions type IIA world-sheet instantons
also contribute to the breaking of the continuous isometry.

Even though the continuous isometry of the Calabi-Yau moduli space is
broken,
we want to argue that our M-theory backgrounds retain a subgroup of
this isometry.
The key difference from the $CY_3\times S^1$ background is that the non-trivial monodromy
has the effect that in the four-dimensional effective action part of the
would-be isometry
(which is indeed an isometry at infinite ${\mathbf P_1}$ volume) is gauged
(see eqs.~\eqref{epsgtr} and \eqref{etagtr}).
Since it is part of a gauge symmetry, consistency requires
that it must persist in the four-dimensional effective action for any
value of the parameters -- and in particular for finite $\mathbf{P_1}$
volume. To reiterate, this must be true even when
 the continuous symmetry is not present for the theory without
the monodromy, or in the five dimensional effective action.

In order to see in slightly more details how this happens
let us first reconsider the computation of the
four-dimensional effective action performed
in section~\ref{sec:KK}. Without the isometry in the Calabi-Yau moduli
space the intersection numbers $\cK_{ijk}$ defined in eq.~\eqref{Khatdfb}
with $\omega_i$ obeying eq.~\eqref{infntsa} are
$z$-dependent and thus vary along the circle. Instead, it is
the intersection numbers $\hat\cK_{ijk}$ defined in \eqref{intno} that
appear in the four-dimensional effective action. Now they no longer
coincide with the $\cK_{ijk}$ as was the case in the presence of a
Calabi-Yau isometry. Nevertheless, if we still require that the monodromy is
evenly distributed along the circle, or in other words if we continue to impose
\eqref{do} for constant $M_i^j$, then eq.~\eqref{conseq} implies
\begin{equation}
  \label{Khcon}
  \M_i^l {{\hat \cK}}_{jkl} + \M_j^l {{\hat \cK}}_{kil} + \M_k^l {{\hat
  \cK}}_{ijl} = 0 \ .
\end{equation}
Thus, the Ansatz \eqref{infntsa} with constant $M_i^j$ implies the
presence of an isometry in the moduli space of $X_7$ even though the
isometry of the fibred Calabi-Yau manifold is broken. The existence of
this isometry can be viewed as a direct
consequence of the gauge symmetry.

The KK-reduction of section \ref{sec:KK} can be repeated, but now
$\hat\cK_{ijk}$ and the metric
defined by \eqref{defg} appear. This metric coincides with the
Calabi--Yau moduli space metric \eqref{defgb} only for infinite $\mathbf{P_1}$,
but differs for finite volume.
Therefore the  resulting four-dimensional effective action  receives
small corrections at finite volume.
However, these corrections
cannot lead to any qualitative changes, since already at large ${\bf P_1}$
volume all fields relevant for the gauging are massive, and the corrections
just shift their precise mass spectrum.

It would  be worthwhile to compute the low energy effective action
more explicitly
and check its consistency with $N=2$ supergravity. Furthermore,
arguments along the lines of
Refs.~\cite{TK-P,AZ} should exist in order to argue that the gauged symmetry
is also protected against the
breaking coming from the world-sheet instantons. We hope to return to
these issues elsewhere.

\section{Heterotic string theory compactified on $K3 \times T^2$ with $T^2$
    fluxes}
\label{hetsect}

In this section we discuss the heterotic string compactified on
$K3 \times T^2$ with the gauge fields having non-trivial flux on the $T^2$.
More specifically we show that the dual background is related to the
M-theory compactification
discussed in the previous section.  We begin by reviewing the
heterotic compactification in sections \ref{genpro}--\ref{N2het},
and we present the details of the duality map in section \ref{mcomparison}.

\subsection{General properties}
\label{genpro}

Consider heterotic string theory compactified on $K3 \times T^2$.
In this subsection we analyze the effect of turning on gauge flux on
the $T^2$ in the low-energy supergravity theory. In particular we want to
show
that turning on the flux breaks the corresponding gauge symmetry, giving
the gauge field a mass
proportional to the flux.

In ten dimensions the spectrum of the heterotic string includes a
2-form field $B$ and a gauge field $A$ with field
strength $F$ (in either
the $Spin(32)$ or the $E_8\times E_8$ gauge group). The 3-form field
strength involves not just the 2-form field, but rather it takes the
form:
\begin{equation}
H^{het} = d B - \frac{\alpha'_{het}}4 \omega_3,
\end{equation}
where $\omega_3$ is the
Chern-Simons form\footnote{There is also a gravitational Chern-Simons term in $H^{het}$, which is of higher
order in the Planck constant and will not play a role in our discussion.}
\begin{equation}
\omega_3 = {\rm tr}(A \wedge dA + \frac{2}{3} A \wedge A \wedge A).
\end{equation}
The ten-dimensional action includes kinetic terms proportional to
\begin{equation}
\left[-\frac{1}{2} |H^{het}|^2 - \frac{\alpha'_{het}}{4} {\rm tr}(F^2) \right].
\end{equation}

Suppose that the compactification to six dimensions on $K3$
breaks the gauge group such that it has a $U(1)^n$ factor, and
consider a background where we turn on a flux for one of the corresponding
$U(1)$ gauge fields $A^{a}$ on the $T^2$ ($a=1,\cdots,n$),
\begin{equation} \label{u1flux}
\int_{T^2} F^{a} \equiv f^{a} \neq 0.
\end{equation}
The six dimensional action includes a term proportional to
\begin{equation} \label{6daction}
\int_{R^4\times T^2} \left[ \big(d B - \frac{\alpha'_{het}}{4} A^{a} \wedge
  F^{a} \big)^2 + \frac{\alpha'_{het}}{2} (F^a)^2 \right],
\end{equation}
such that the four dimensional action expanded around the flux
background \eqref{u1flux} includes a term proportional to
\begin{equation} \label{4daction}
\int_{R^4} \left[ \big( d b - \frac{\alpha'_{het}}{4} f^{a} A^{a} \big)^2 +
  \frac{\alpha'_{het} V(T^2)^2}{2} (F^a)^2 \right],
\end{equation}
where $b$ is the scalar field arising from $\int_{T^2} B$, and $V(T^2)$ is the
volume of the $T^2$.
Naively the first term is not gauge-invariant, but in fact the gauge
transformation (already in ten dimensions) acts also on the
2-form field, and this transformation in four dimensions takes
the form $A^{a} \to A^{a} + d \Lambda^{a}$, $b \to b + \frac{\alpha'_{het}}{4} f^{a} \Lambda^{a}$ such
that \eqref{4daction} is gauge-invariant.

Both from the form of \eqref{4daction} and from the form of
the gauge transformation, we see that the $U(1)$ gauge symmetry
is broken, since it acts non-linearly on the scalar field $b$.
The gauge field acquires a mass proportional to $f^{a}$, and swallows
the scalar field $b$ by the Higgs mechanism. Using \eqref{4daction} we
see that the mass squared of the gauge field is proportional to
$\alpha'_{het} f_a^2 / V(T^2)^2$. In the action we
wrote here we set many fields to zero, the full results may
be found in \cite{LM1}.

In the previous section we saw that a similar Higgs mechanism in M-theory arises
from the non-trivial fibration structure over the M-theory circle. In the
following we argue why it is indeed necessary to go to the M-theory
description on the dual type IIA side when we add the heterotic fluxes, 
and afterwards we make the correspondence between the M-theory and
the heterotic Higgsing more precise.

\subsection{Mapping the masses}
\label{mapmas}

In order to map the Higgs mechanism described above to the type IIA
side, we need to compute the mass of the massive vector, and describe
it in the language of the type IIA string theory.

Let us first recall the mapping in the absence of fluxes between the
heterotic string and the type IIA string. On the heterotic side,
the $K3$ manifold is taken to be a fibration of $T^2_f$ over
some base $B$. On the type IIA side we have a Calabi-Yau manifold
which is a fibration of some ${\tilde K3}$ over $B$ (where we used
fiber-wise the duality between the heterotic string theory on $T^4$
and the type IIA string theory on ${\tilde K3}$).

The relations between the parameters of the two theories are (denoting
the volume of a cycle by $V$, and not writing down all the numerical
constants) :

The mapping of the four dimensional Planck scales gives
\begin{equation}
V(K3)V(T^2)/g_h^2l_h^8=V({\tilde K3})V(B)/g^2_{II}l^8_{II}.
\end{equation}

For the mapping of the type IIA string to a wrapped heterotic five-brane we
have
\begin{equation}
V(T^2)V(T^2_f)/g^2_hl^6_h= 1/l^2_{II}.
\end{equation}

The mapping of the heterotic string to a wrapped NS5-brane yields
\begin{equation}
1/l^2_h=V({\tilde K3})/g^2_{II}l^6_{II}.
\end{equation}

Finally, the integral of the heterotic $B$-field on the $T^2$ maps
to the integral of the type IIA $B$-field on some 2-cycle $W$ in
$\tilde K3$, leading to
\begin{equation}
V(T^2)/l^2_h=V(W)/l^2_{II}.
\end{equation}

Above we found that on the heterotic side
the mass of the vector field that becomes massive after we turn on
the flux is
\begin{equation}
m^2=(f^{a})^2 l_h^2 / V(T^2)^2.
\end{equation}
Translating this into type IIA string theory using the equations above,
we find that the mass can be written as
\begin{equation}
m^2 = (f^{a})^2 V({\tilde K3})/ (V(W)^2 g_{II}^2 l_{II}^2).
\end{equation}

In particular, it involves a negative power of the type IIA string
coupling, implying that it is not a perturbative state on the type
IIA side. Rather, since its mass is proportional to the D0-brane
mass $M_{D0}\simeq 1 / g_{II} l_{II}$, it involves when lifted to
M-theory some non-trivial momentum on the M-theory circle. Thus, we
cannot describe this flux purely in the language of type IIA
supergravity (the massive gauge field is too massive to be included
in the low energy IIA description). The dual configuration must involve,
when lifted to M-theory, non-trivial dependence on the M-theory circle.

\subsection{The flux as a monodromy}
\label{fluxmon}

We claim that the correct description of this flux on the type IIA
side is given by the non-trivial fibration of the Calabi-Yau over the
M-theory circle, described in the previous section. In order to make
this identification more precise, let us move up one dimension, and
consider the heterotic string theory on $K3\times S^1$, which is dual
to M-theory on a Calabi-Yau manifold (this is simply the limit of the
duality discussed in the previous subsection, when one of the heterotic
circles is taken to be large). We will call the coordinate on this
circle $x^5$, and denote the coordinate on the additional circle which
we use to go down to four dimensions by $x^4$ (this may be identified with
the $z$ coordinate which we used in section \ref{sec:geo}).

In the $K3\times S^1$
compactification, each ten-dimensional gauge field
$A_{\mu}^{a}$ leads to a scalar field
$A_5^{a}$. One way to describe the flux that we are interested in is
by taking this scalar field to have a non-trivial monodromy around the
additional circle in the $x^4$ direction,
\begin{equation}
\label{a5shift}
A_5^{a} = c f^{a} x^4 \Rightarrow A_5^{a}(x^4+2\pi R_4) \simeq
A_5^{a}(x^4) + 2\pi c f^{a} R_4
\end{equation}
for some constant $c$.
Note that the low-energy supergravity is invariant under any shift in
the scalar field $A_5^{a}$; however, in the full heterotic string theory,
due to the presence of charged states carrying momentum on the $x^5$ circle,
there is only a discrete periodicity of the field $A_5^{a}$. Equation
\eqref{a5shift} may be interpreted as saying that when we go around the $x^4$
circle, $A_5^{a}$ comes back to itself up to a shift by an integer multiple of its
period (proportional to $f^a$).

In this language, we can think of the flux as a special case of a monodromy in
the T-duality group. Recall that the heterotic string theory on $K3\times S^1$
has $n$ $U(1)$ vector fields $A_{\mu}^{a}$ coming from the ten-dimensional
gauge group, and three additional vector fields coming from $g_{\mu 5}$,
$B_{\mu 5}$ and the dual of $B_{\mu \nu}$. One combination of the three latter
fields is in the graviton multiplet, while the other $n_V^{(5)}=n+2$ fields
are in vector multiplets.  Each of the vector multiplets contains a real
scalar field; these $n_V^{(5)}$ fields are $A_5^{a}$, the radius of the $x^5$
circle, and the heterotic dilaton, and they span the manifold
\cite{Gunaydin:1984ak,AFT} $SO(1,n_V^{(5)}-1)/SO(n_V^{(5)}-1)\times {\bf R}$.
The low-energy supergravity action is invariant under an
$SO(1,n_V^{(5)}-1)\times SO(1,1)$ symmetry, where the first factor rotates the
scalars (and all the vector fields except for the dual of $B_{\mu \nu}$),
while the second factor shifts the dilaton.  In the full heterotic string
theory, only an $SO(1,n_V^{(5)}-1,{\bf Z})$ subgroup of this group is an exact
symmetry -- this is the T-duality group of the heterotic string on a circle.
This group includes in particular the shifts in $A_5^{a}$ described in the
previous paragraph. Thus, these shifts are a special case of a general
$SO(1,n_V^{(5)}-1,{\bf Z})$ monodromy, where as we go around the circle the
theory comes back to itself up to some $SO(1,n_V^{(5)}-1,{\bf Z})$
transformation.

It is now clear, that in order to map the flux to the M-theory side, we need
to consider backgrounds in which M-theory on a Calabi-Yau comes back to itself
(as we go around the circle) up to some element of $SO(1,n_V^{(5)}-1,{\bf Z})$. These are
precisely the backgrounds we considered in the previous section, so we claim
that these are the correct type II duals of the heterotic compactification
with flux. In the next two subsections we will check this proposal in detail,
by mapping the four dimensional effective actions of the two theories.

\subsection{The low-energy effective action}
\label{N2het}

Let us briefly recall the low energy effective action for heterotic string
compactifications on $K3 \times T^2$ with non-trivial background fluxes, which
was derived in \cite{LM1}.  In the spirit of the present paper we only focus
on the vector multiplets and only review the low energy theory for fluxes of
the gauge fields on $T^2$, as they lead to a non-Abelian gauge group in the
effective four-dimensional theory. The main features of the ungauged theory are
summarized in appendix \ref{vshet}.

The $n_v=n+3$ four dimensional heterotic vector multiplets include the complex
scalar fields $x^i = (s, u, t, n^a),~ a=4,\ldots, n_v$ which span the
symmetric space \eqref{vmshet}, where $s$ denotes the dilaton/axion, $t$ and
$u$ are the $T^2$ moduli and $n^a$ denotes the scalars arising from the Wilson
lines of the original heterotic gauge fields in the $T^2$ directions. The
latter combine with the four-dimensional gauge fields $A^a$ which also
originate from the ten-dimensional heterotic gauge fields. From the metric and
the $B$-field we obtain four Kaluza-Klein gauge bosons $A^0,\ldots, A^3$ which
play the role of the graviphoton and the superpartners of $s,t$ and
$u$.\footnote{The details can be found in reference \cite{LM1}.}
In the absence
of fluxes the gauge group is the Abelian group $[U(1)]^{(n_v+1)}$.

When we turn on background fluxes of the form
\begin{equation}
  \label{hetflux}
  \int_{T^2} F^a = f^a \; ,
\end{equation}
the four dimensional gauge group becomes non-Abelian (in the sense
that different gauge transformations no longer commute), as in the
general gauged supergravities discussed in the appendix. Note that
this non-Abelian symmetry has nothing to do with the original $E_8
\times E_8$ or $SO(32)$ gauge symmetry in ten dimensions; it
involves only fields in the Cartan subgroup of the original gauge
group.

The action  computed in \cite{LM1} is\footnote{Compared to \cite{LM1} we have rescaled the metric by a factor $1/2$ and
the gauge fields by a factor $1/\sqrt 2$ in order to agree
  with the conventions we use in type IIA compactifications.}
\begin{equation}
  \label{Sh4}
  S_{\mathrm{het}} = \int \left[ \tfrac12 R *1 + \tfrac14 I_{IJ} F^I \wedge *
  F^J + \tfrac14 R_{IJ} F^I \wedge F^J - g_{ij} D x^i \wedge * D
  \bar x^{\bar \jmath} - V \right] \ ,
\end{equation}
which slightly differs from the action given in \eqref{agsg}. The
point is that from the heterotic viewpoint a different symplectic
basis is more natural. More precisely, the gauge field $A_1$ is
dualized relative to the formalism used in the appendix, which is
the one we use for M-theory. In this basis the prepotential $\cF$
does not exist but its derivatives are well defined
\cite{CCDF,dWKLL,AFGNT}. So let us carefully go through the terms.

The non-trivial covariant derivatives in \eqref{Sh4} when we turn on the fluxes are given
by
\begin{equation}
  \label{cdh}
  \begin{aligned}
  D t = & ~ \partial t - \sqrt 2\, n^a f^a A^1 + f^a A^a \; ,\\
  D n^a = &  ~ \partial n^a - \tfrac{1}{\sqrt 2}\, f^a (A^0 + u A^1) \; ,
  \end{aligned}
\end{equation}
which, using \eqref{gaugeco}, corresponds to the Killing vectors
\begin{equation}
  \label{kvh}
  k_0  =  \tfrac{1}{\sqrt 2}\, f^a \partial_a \; , \qquad
  k_1  =  \tfrac{1}{\sqrt 2}\, f^a u\, \partial_a + \sqrt 2\, n^a f^a \partial_t \; ,
  \qquad
  k_a  =  - f^a \partial_t \ .
\end{equation}
Finally, the metric $g_{ij}$ in \eqref{Sh4} is special K\"ahler and can be
derived from \eqref{kpothet}.

As explained in appendix \ref{vshet}, the gauge couplings $I_{IJ}, R_{IJ}$,
which are given in \eqref{gkfhet}, cannot be derived directly from
\eqref{Ndef}.
In the ungauged case ($f^a=0$) one needs to perform an electric-magnetic
duality transformation on the symplectic vector $X^I, \cF_I$ given by
$X^1 \to -\cF_1$ and $\cF_1 \to X^1$. Using \eqref{nchange} this
transforms the gauge couplings $I_{IJ}, R_{IJ}$ into a form consistent with
\eqref{asg} and \eqref{Ndef} while the K\"ahler potential is left
invariant.
For the gauged case ($f^a\neq0$) this transformation is not
straightforward and generates precisely a term of the form \eqref{dL}
as we will see in the next subsection.

The non-Abelian field strengths in the
heterotic basis are given by
\begin{eqnarray}
  \label{hetfs}
  F^0 & = & d A^0 \; , \nn \\
  F^1 & = & d A^1 \; , \nn \\
  F^2 & = & d A^2 +  f^a A^a \wedge A^1 \; , \\
  F^3 & = & d A^3 -  f^a A^a \wedge A^0 \; ,\nn  \\
  F^a & = & d A^a -  f^a A^0 \wedge A^1 \; . \nn
\end{eqnarray}
The equations can be understood as follows: recall that (when we do
not turn on any non-trivial fields) $A^0$ and $A^1$ are linear
combinations of $g_{\mu 4}$ and $g_{\mu 5}$, while $A^2$ and $A^3$
are linear combinations of $B_{\mu 4}$ and $B_{\mu 5}$. The
non-Abelian terms in $F^2$ and $F^3$ follow from \eqref{6daction}
when including off-diagonal metric elements in the contractions. The
non-Abelian term in $F^a$ arises just from off-diagonal contractions
in the standard six dimensional kinetic term of $F^a$. By comparing
with \eqref{strconA} we see that
the non-vanishing structure constants are
\begin{equation}
  \label{fvh}
  f^2_{a1} = - f^3_{a0} = f^a_{01} =  f^a\ .
\end{equation}

Note that there is a slight subtlety when one takes the Killing
vectors as given in \eqref{kvh} and checks the consistency of
\eqref{fvh} with \eqref{commutator}. The reason is that the structure
constants \eqref{fvh} correspond to a Lie algebra generated by
$(T_0, ~T_1,~ T_2,~ T_3,~ T_a)$ obeying
\begin{equation}
  \label{alg}
  \left[ T_0 , T_1 \right] = f^a T_a \; , \qquad \left[ T_0 , T_a \right] =
  f^a T_3 \; , \qquad \left[ T_a , T_1 \right] = f^a T_2 \; ,
\end{equation}
with all the other commutators vanishing.
We see that $T_2$ and $T_3$ are central elements of the algebra and
therefore can consistently be set to zero. This is precisely what
happened in our case in that the Killing vectors
$k_2$ and $k_3$ are vanishing in \eqref{cdh}, and therefore the
last two commutators in \eqref{alg} are zero even though the
corresponding structure constants are non-zero. This situation is
encountered frequently in gauged supergravities, see for
example~\cite{hull1,GRZ}.\footnote{We thank Marco Zagermann for educating us
  on this subject and the referee of this paper for pointing out that
  for $T_2=T_3=0$ \eqref{alg} is also a nilpotent Lie algebra
  \cite{Turin}.}

Finally, the potential in the action \eqref{Sh4} is given by the
standard formula \eqref{pot} with the Killing vectors \eqref{kvh} inserted.

\subsection{Comparison to M-theory}
\label{mcomparison}

In this section we wish to compare the heterotic flux
compactification derived in the previous subsections, with the
M-theory compactification of the previous section. For this we have to
remember
that in the ungauged case the map between heterotic and type IIA theories
involves the non-trivial symplectic rotation \eqref{srot}. On the gauge fields
this translates into the map
\begin{eqnarray}
  \label{Aid}
  A^0_{\mathrm{het}} & \equiv & -A^2_{\mathrm{IIA}} \; , \nn \\
  A^1_{\mathrm{het}} & \equiv & A^0_{\mathrm{IIA}} \;  ,\nn \\
  A^2_{\mathrm{het}} & \equiv & A^3_{\mathrm{IIA}} \; , \\
  A^3_{\mathrm{het}} & \equiv & \tilde A^1_{\mathrm{IIA}} \; , \nn \\
  A^a_{\mathrm{het}} & \equiv & \sqrt 2 A^a_{\mathrm{IIA}} \; , \nn
\end{eqnarray}
where $\tilde A^1$ denotes the electric-magnetic dual of the vector field $A^1$ which appears
in the type IIA compactification.

In order to compare
the low-energy effective actions, we need to insert the $\M_i^j$
into eq.~\eqref{kvA} and compare the resulting covariant derivatives and
Killing vectors to the heterotic side as given in \eqref{kvh}. We
immediately see that there is no perfect match between all the M-theory
parameters and the heterotic fluxes that we discussed thus far, and
we will return to this point later.

However, let us first see for which subset of the M-theory torsion
parameters, the heterotic flux can be recovered. Indeed, choosing
\begin{eqnarray}
  \label{hetpar}
  m_2=  m_3= m_a = m_b^a = 0\ ,
\end{eqnarray}
and leaving only $\tilde m_a\neq 0$ in
eq.~\eqref{kvA} results in the non-trivial covariant derivatives
\begin{equation}
  \label{cdIIA}
  \begin{aligned}
    D_\mu x^3 = \partial_\mu x^3 + \tilde m_a(x^a A^0_\mu - A^a_\mu) \; , \\
    D_\mu x^a = \partial_\mu x^a + \tfrac12 \tilde m_a (x^2 A^0_\mu  - A^2_\mu
    ) \; ,
  \end{aligned}
\end{equation}
or  equivalently the Killing vectors
\begin{eqnarray}
  \label{kspecial}
  k_0^3  = - x^a \tilde m_a\ ,\qquad  k_0^a = - \tfrac12 x^2 \tilde m_a\
  ,\qquad 2k_2^a = k_a^3=\tilde m_a \ .
\end{eqnarray}
Comparison with eq.~\eqref{kvh} together with the identifications
\eqref{Xid} and \eqref{Aid} shows a perfect match if we identify
\begin{equation}
  \label{fluxid}
  \tilde m_a |_{\mathrm{IIA}} = - \sqrt2 f^a |_{\mathrm{heterotic}} \; .
\end{equation}

We can similarly compare the field strengths. Inserting eq.~\eqref{hetpar}
into \eqref{cd} we arrive at
\begin{equation}\begin{aligned}
  \label{FIIA}
  F^3 & =  d A^3 + \tilde m_a A^0 \wedge A^a \; , \\
  F^a & =  d A^a - \tfrac12 \tilde m_a A^0 \wedge A^2 \; .
\end{aligned}\end{equation}
Comparing with eq.~\eqref{hetfs} using eqs.~\eqref{fluxid} and \eqref{Aid} we
see that the field strengths $F^3$ and $F^a$ above precisely correspond to
$F^2$ and $F^a$ on the heterotic side. However, $F^1$ on the type IIA/M-theory
side is Abelian while its correspondent (via \eqref{Aid}), $F^3$, on the
heterotic side is non-Abelian. On the other hand  the M-theory
side has an additional term (the last term in \eqref{S4}) in the
low energy effective action. The reason for this mismatch is the fact
that the two actions are computed in different symplectic
frames. In the ungauged case (i.e.\ for  $\tilde m_a=0$) one easily
identifies a symplectic rotation which connects the two frames. In the
gauged case (i.e.\ for  $\tilde m_a\neq 0$) this is less
straightforward and will occupy us for the rest of this
section.\footnote{The  following discussion  should be straightforward
  in the framework of gauged supergravity as given in
  \cite{dWST}.}

Let us first recall that the presence of the last
term in eq.~\eqref{S4} was due to the fact that the prepotential
\eqref{prep} was not invariant under the gauge transformation
\eqref{etagtr}. However in the heterotic frame all terms in
eq.~\eqref{Sh4} are invariant  and this term is absent.
For the choice of parameters \eqref{hetpar} the last term in
eq.~\eqref{S4} becomes (up to a total derivative)
\begin{equation}
  \label{AAF}
  - \tfrac12 \tilde m_a A^2 \wedge A^a \wedge dA^1 \; .
\end{equation}
In order to have the two sides match we have to exchange the gauge
field $A^1$ with its magnetic dual.\footnote{Recall that already in
 six dimensions, the duality between heterotic string theory on $T^4$
 and type IIA string theory on $K3$ involves a dualization of the
 2-form field.} This is indeed possible as the gauge field $A^1$ appears only
via its (Abelian) field strength $F^1=d A^1$ as can be seen from
eqs.~\eqref{S4} and \eqref{AAF}. The easiest way to see how to do
the dualization is to add a Lagrange multiplier $ - \tfrac12 F^1 \wedge d
{\tilde A}_1$ which enforces the Bianchi identity of $F^1$, and ${\tilde A}^1$
will become the magnetic dual of the gauge field $A^1$.
The equation of motion for $F^1$ then reads
\begin{equation}
  \label{Feom}
  \tfrac12\, \mathrm{Im} \cN_{1J} * F^J + \tfrac12\, \mathrm{Re} \cN_{1J} F^J
  - \tfrac12 \tilde m^a A^2 \wedge A^a - \tfrac12 d {\tilde A}_1 = 0 \; .
\end{equation}
Defining now the magnetic dual field strength $G_1$ as
\begin{equation}
  \label{G1}
  G_1 = d {\tilde A}_1 + \tilde m^a A^2 \wedge A^a \; ,
\end{equation}
the equation of motion for $F^1$ becomes
\begin{equation}
  \label{mgfs}
  \tfrac12\, G_1 ~ = ~
  \tfrac12\, \IM \cN_{1J} * F^J + \tfrac12\, \RE \cN_{1J} F^J
  ~ \equiv ~ \frac{\partial \cL_{N=2}}{\partial F^1}\ \; ,
\end{equation}
where $\cL_{N=2}$ denotes the generic $N=2$ Lagrangian \eqref{agsg}. This
equation is precisely the definition of magnetic dual field strength in $N=2$
supergravities \eqref{eom} and from here on we can apply the general
dualization procedure and transform the matrix of gauge couplings $\cN$ as in
\eqref{nchange} with the matrices $U$, $V$, $Z$ and $W$ chosen such that $F^1
\to G_1$ in \eqref{FGdual}.

Clearly, now $G_1$ defined in eq.~\eqref{G1} can be mapped to the heterotic
field strength $F^3$ from eq.~\eqref{hetfs}, via eqs.~\eqref{Aid} and
\eqref{fluxid}. 
This ends the proof that the low energy theories
obtained from compactifying heterotic strings on $K3 \times T^2$ with
fluxes turned on along $T^2$ and from compactifying M-theory on a
seven-dimensional manifold with $SU(3)$ structure with only the fluxes
$\tilde m_a$ non-vanishing, are indeed the same.

So far we discussed the duality for the parameter choice \eqref{hetpar}.
However, our discussion in the previous section makes it clear that all the
parameters $M^i_j$ on the M-theory side, which give rise to
consistent backgrounds in the full M-theory\footnote{Namely, $e^M$ must be
a member of the discrete U-duality group.}, correspond to
$SO(1,n_V^{(5)}-1,{\bf Z})$ monodromies, and they can be described
by such monodromies on the heterotic side as well. The specific
monodromy we discussed above is simple on the heterotic side since
it does not involve the metric, but it is just a shift of the Wilson
lines $A_5^{(i)}$ around the torus $T^n$ that they live on.
Monodromies in an $SO(n,{\bf Z})$ subgroup of $SO(1,n_V^{(5)}-1,{\bf
Z})$ may be identified as $SL(n,{\bf Z})$ transformations on this
torus, which mix the various gauge fields and scalars; these were
denoted by $M^a_b$ above. Generic monodromies (involving $m_2$,
$m_3$ and $m_a$) do not have a purely geometrical description
\cite{LMM}. For
instance, the $m_a$ parameters are related by a T-duality (inverting
the radius of one of the circles) to the $\tilde m_a$ parameters, so
they may be viewed as having a variation of the heterotic gauge
fields $A^a$ (similar to \eqref{a5shift}) along the T-dual circle.
However, this ``T-dual flux'' does not have a geometrical description in
the original heterotic language.
Finally note that a background with $m_2 + m_3 \ne 0$ is not consistent as it
involves a twist with an element of $SO(1,1, {\bf Z})$ which is not part of
the U-duality group in five dimensions. This can also be seen from the
heterotic side as it would make the heterotic dilaton charged, which has not
been observed so far in perturbation theory.

\section{Conclusions}

In this paper we studied M-theory compactifications on seven-dimensional
manifolds with $SU(3)$-structure. Specifically we considered a class of such
manifolds which can be seen as Calabi--Yau threefolds fibred over a circle.
The fibration structure is determined by a specific twist of the second
cohomology of the Calabi--Yau as we go around the circle. The consistency of
the procedure requires that a discrete isometry in the Calabi--Yau moduli
space exists (which is an element of the U-duality group of M-theory
compactified on the Calabi-Yau manifold). This is guaranteed for $K3$-fibered
Calabi--Yau manifolds which correspond to backgrounds that are dual to the
heterotic string compactified on $K3\times T^2$.

Since in such compactifications the second cohomology of the Calabi--Yau
manifold governs the vector multiplet sector, the twisting leads to a gauged
supergravity where a subset of the isometries of the vector multiplet moduli
space are promoted to local gauge symmetries. A novel feature is that the
K\"ahler moduli are charged, and not only their axionic superpartners as it
usually happens in $N=2$ string compactifications. Moreover this gauging turns
out to be non-Abelian which so far had not been obtained in (smooth)
compactifications of type IIA string theory or M-theory.

The fact that this gauging should exist is expected from the heterotic -- type
IIA duality. In heterotic $N=2$ backgrounds arising from
$K3\times T^2$ compactifications with specific background fluxes
only the vector multiplets get charged and
the potential has no dependence on the hypermultiplets.
However, viewed from the dual type IIA perspective, the masses of
the vector fields contain negative powers of the type IIA
string coupling. Therefore, in order to consistently keep such states in the
effective theory and at the same time ignore the KK states, one has to make
sure that the type IIA string coupling is large relative to the size of the
Calabi--Yau manifold. This  forced us into the M-theory regime, and indeed
the dual of the heterotic backgrounds were found
among the M-theory backgrounds described above.

The general twisted compactification on the M-theory side contains additional
parameters which do not map to fluxes on the heterotic side. However, since we
can interpret all such compactifications as twists of the five dimensional
theory (obtained from M-theory on the Calabi-Yau, or equivalently from the
heterotic string theory on $K3\times S^1$) by an element of the heterotic
T-duality group, they can all be described as T-folds on the heterotic side.
It would be interesting to study these backgrounds further; work along these
lines is in progress \cite{LMM}.

\vskip 1cm

\noindent
\textbf{\large Acknowledgments}

This work was supported by G.I.F., the German-Israeli Foundation for Scientific Research
and Development.
The work of OA and MB was  supported in part by the
Israel-U.S. Binational Science Foundation, by a center of excellence supported by the Israel Science Foundation
(grant number 1468/06), by a grant (DIP H52) of the German Israel Project Cooperation, by the European network MRTN-CT-2004-512194, by Minerva, and by the Einstein-Minerva Center for
Theoretical Physics.
The work of AM was supported by
the FP6 Marie Curie Research Training Networks,
the  European Union 6th framework program MRTN-CT-2004-503069 ``Quest for
unification", MRTN-CT-2004-005104 ``ForcesUniverse",
MRTN-CT-2006-035863 ``UniverseNet", and
the Deutsche Forschungsgemeinschaft (DFG) in the
SFB-Transregio 33 "The Dark Universe".
The work of JL was supported by
the  European Union 6th framework program MRTN-CT-2004-503069 ``Quest for
unification", and
the Deutsche Forschungsgemeinschaft (DFG) in the SFB 676 ``Particles,
Strings and the Early Universe''.

OA and MB would like to thank Albion Lawrence for useful discussions.  JL and
AM thank Ron Reid-Edwards, Thomas Grimm, Danny Martinez, Eran Palti, Bastiaan
Spanjaard, Daniel Waldram and Marco Zagermann for helpful conversations. JL
thanks Chris Hull and the Institute for Mathematical Sciences, Imperial
College London, for financial support and the kind hospitality during part of
this work.

\vskip1cm
\appendix


\section{Vector multiplets coupled to $N=2$ supergravity}
\label{sg}

This appendix is a short review of $N=2$
supergravity in four dimensions \cite{dWvP,N=2review}.  A generic spectrum contains the
gravitational multiplet, $n_V$ vector multiplets, $n_H$
hypermultiplets and $n_T$ vector multiplets.
In this paper we are
interested only in the vector multiplet sector and therefore we discard  the
hyper- and tensor-multiplets.

The vector multiplets contain $n_V$ complex scalars $x^i,
i=1,\ldots,n_V$,
which span  a special K\"ahler manifold $\cM_V$.
This implies  that the K\"ahler potential $K$ is not an arbitrary
real function but is determined in terms of a holomorphic prepotential $\cF$
according to \cite{dWvP}
\begin{equation}
  \label{Kspecial}
  K=-\ln\Big[i \bar{X}^{I} (\bar x) \cF_{I}(X)
  - i X^{I}  (x)\bar{\cF}_{I}(\bar{X})\Big] \ .
\end{equation}
The $X^{I}, I=0,\ldots, n_V$ are $(n_V+1)$ holomorphic functions
of the scalars
$x^i$, and  $\cF_{I}$ abbreviates the derivative, i.e.  $\cF_{I}\equiv
\frac{\partial \cF(X)}{\partial X^{I}} $. Furthermore $\cF(X)$ is a homogeneous
function  of degree $2$ in $X^{I}$, i.e.\ $X^{I} \cF_{I}=2 \cF$.

The bosonic part of the (ungauged) $N=2$ action for vector multiplets is given by
\begin{equation}
\label{asg}
  S =  \int \Big[ \frac12 R ^* {\bf 1} - g_{i\bar\jmath} dx^i \wedge * d
  {\bar x}^{\bar\jmath} 
  + \frac{1}{4}\, \IM \cN_{IJ} F^I\wedge * F^{J}
  + \frac{1}{4} \, \RE \cN_{IJ} F^I \wedge F^J
  \Big] \ ,
\end{equation}
where $g_{i\bar\jmath}=\partial_i\partial_{\bar\jmath} K$. In the
ungauged case the field strengths are Abelian, $F^I = dA^I$,
and the matrix of gauge couplings is given by
\begin{equation}
  \label{Ndef}
  {\cal N}_{IJ} = \bar \cF_{IJ} +2i\ \frac{\mbox{Im} \cF_{IK}\mbox{Im}
    \cF_{JL} X^K X^L}{\mbox{Im} \cF_{LK}  X^K X^L} \ .
\end{equation}

The equations of motion of the action (\ref{asg}) are invariant under
generalized electric-magnetic duality transformations.
{}From (\ref{asg}) one derives the equations of motion
\begin{equation}
  \label{eom}
  \frac{\partial \cL}{\partial A^I} =  \tfrac12 d {G}_I = 0 \ , \qquad
  {G}_I \equiv 2 \frac{\partial \cL}{\partial F^I} = \RE \cN_{IJ} {F}^J  + \IM
  \cN_{IJ} *F^J \, ,
\end{equation}
while the Bianchi identities read
\begin{equation}
d {F}^I = 0\ .
\end{equation}
These equations are invariant under the generalized duality rotations\footnote{
This is often stated in terms of the self-dual and anti-self-dual part of the
field strength $F^{\pm J}$ and the dual quantities
$G^+_{I}\equiv{\cal N}_{IJ}F^{+J}\,,\
G^-_{I} \equiv\bar{\cal N}_{IJ}F^{-J}$.}
\begin{eqnarray}
  \label{FGdual}
  F^{I}&\to&
  U^I{}_J\, F^{J}+Z^{IJ}\,G_{ J}\ ,\nn\\
  G_I&\to& V_I{}^J\,G_{J}+W_{IJ}\,F^{J}\ ,
\end{eqnarray}
where $U$, $V$, $W$ and $Z$ are constant, real,  $(n_V+1)\times(n_V+1)$
matrices which obey
\begin{eqnarray}
  \label{spc2}
  U^{\rm T} V- W^{\rm T} Z &=& V^{\rm T}U - Z^{\rm T}W =
  {\bf 1}\, ,\nn\\
  U^{\rm T}W = W^{\rm T}U\,, && \quad Z^{\rm T}V= V^{\rm T}Z\ .
\end{eqnarray}
Together they form the $(2n_V+2)\times(2n_V+2)$ symplectic matrix
\begin{equation}
  \label{uvzwg}
  {\cal O}\ = \left(
    \begin{array}{cc}
      U & Z \\
      [1mm] W & V
    \end{array}
  \right) \, .
\end{equation}
Thus $(F^I,G_I)$ form
a $(2n_V+2)$ symplectic vector. Similarly $(X^I,\cF_I)$ enjoy the same
transformation properties and transform as a
symplectic vector under (\ref{FGdual}). The K\"ahler  potential
(\ref{Kspecial}) is invariant under this symplectic transformation,
while  the matrix $\cN$ transforms according to
\begin{equation}
  \label{nchange}
  \cN \to (V \cN+ W) \,(U+ Z \cN)^{-1} \,.
\end{equation}

The isometries of the scalar manifold $\cM_V$
are global invariances of the scalar field sector, which can be ``gauged'' by
mixing them with the
local symmetries.  These isometries are generated by holomorphic
Killing vectors $k_I^i(x)$ via
\begin{equation}
  \label{kdef}
  \delta x^i \ = \ \Lambda^I k_I^i(x) \  .
\end{equation}
The $k_I^i(x)$ satisfy the Killing equation which in $N=2$ supergravity can be
solved in terms of a Killing prepotential $P_I$
\begin{equation}\label{Pdef}
  k_I^i(x) = g^{i\bar j} \partial_{\bar j} P_I\; .
\end{equation}
Gauging the isometries (\ref{kdef}) requires the replacement of
ordinary derivatives by covariant derivatives in the action \eqref{asg}
\begin{equation}
  \label{gaugeco}
  \partial_\mu  x^i \to {D}_\mu  x^i = \partial_\mu  x^i - k_I^i A_\mu^I\ ,
\end{equation}
and the field strengths take the form
\begin{eqnarray}
  \label{strconA}
F^I = d A^I + f^I_{JK}A^J\wedge A^K \ .
\end{eqnarray}
Consistency requires
\begin{equation}
  \label{commutator}
\big[k_I, k_J \big]\ =\ f_{IJ}^L\ k_L\; ,
\end{equation}
where $k_I = k_I^j\partial_j$. Furthermore the potential
\begin{equation}
  \label{pot}
  V  = 2 e^K  X^I\bar X^J g_{\bar \imath j}\, k_I^{\bar\imath} k_J^j
\end{equation}
has to be added to the action in order to preserve
supersymmetry.\footnote{Note the factor 2 in front of the potential compared
  to \cite{N=2review} which comes from the different normalization which we
  use in the action \eqref{agsg}.}
The bosonic part of the action of gauged $N=2$ supergravity is
then given by
\begin{equation}
  \label{agsg}
  S =  \int \Big[ \frac12 R ^* {\bf 1} - g_{i{\bar\jmath}} Dx^i \wedge * D
  {\bar x}^{\bar\jmath} 
  + \frac{1}{4}\, \IM \cN_{IJ} F^I\wedge * F^{J}
  + \frac{1}{4} \, \RE\cN_{IJ} F^I \wedge F^J -V
  \Big] \ .
\end{equation}
The symplectic invariance of the ungauged theory is generically broken since
the action now explicitly depends on the gauge potentials $A^I$ through the
covariant derivatives $Dx^i$ and the non-Abelian field strengths $F^I$.

There is yet a further generalization of the above setup which was discussed
in \cite{dWvP}. The isometries considered above need not leave  the
prepotential $\cF$ invariant. For example, consider an isometry which
leads to a change in the prepotential of the type
\begin{equation}
  \label{dF}
  \delta \cF = \Lambda^I C_{IJK} X^J X^K \; ,
\end{equation}
for some real parameters $C_{IJK}$. Obviously, the imaginary
part of the second derivative of this variation vanishes. From its
definition  \eqref{Ndef} we see that the
imaginary part of the gauge coupling matrix $\mathrm{Im} \cN$ is left
invariant. $\mathrm{Re}\cN$ changes however, and so the action as defined
in \eqref{agsg} is not invariant. In order to restore gauge invariance
the following term has to be added to the action  \cite{dWvP}
\begin{equation}
  \label{dL}
 S \to S+  \int \tfrac13 C_{IJK} A^I \wedge A^J \wedge (d A^K - \frac38 f_{LM}^K
  A^L \wedge A^M) \; .
\end{equation}
%

\section{The vector multiplet sector of heterotic string compactifications
  on $K3 \times T^2$}
\label{vshet}

In this appendix we review the structure of the vector multiplet sector of
heterotic strings compactified on $K3 \times T^2$, following \cite{LM1}. For this setup, the vector
multiplet sector is directly connected to the $T^2$ part of the
compactification and the $K3$ factor only breaks supersymmetry and may reduce
the total number of vector multiplets. Therefore, for our purposes studying
the $T^2$ step will be enough. The initial non-Abelian gauge symmetry of the
heterotic string is in general broken spontaneously to the maximal Abelian
subgroup and therefore we consider the resulting theory to be $N=2$
supergravity coupled to an arbitrary number $n_v$ of Abelian vector
multiplets.

The vector fields in the vector multiplets have two origins: first
they can come from gauge fields in ten dimensions (and their number is
arbitrary) and second they arise as KK vector fields on the torus. In the
last class we have precisely four vector fields, two from the internal
components of the metric -- which we denote $A^0$ and $A^1$ -- and two from the
$B$-field -- which we denote $A^2$ and $A^3$. One of these vector fields, or
some combination of them will be the graviphoton, while the rest will sit in
vector multiplets. The vector fields from the first class we denote as $A^a$
($a=4, \ldots , n_v$), and they are all part of vector multiplets.

The scalar fields in the vector multiplets span the coset space
\begin{equation}
  \label{vmshet}
  \cM_V = \frac{SU(1,1)}{U(1)} \otimes \frac{SO(2,n_v-1)}{SO(2) \times
  SO(n_v-1)} \; .
\end{equation}
The factor $SU(1,1)/U(1)$ corresponds to the dilaton and its
superpartner, the axion dual to the four-dimensional $B$-field, while the
second factor describes the scalar fields coming from the $T^2$ moduli
(including the internal $B$-field) and from the internal components of the
ten-dimensional gauge fields. These fields combine into the complex scalar fields
$x^i = (s, u, t, n^a), ~ a=4, \ldots , n_v$, with $s$ being the heterotic
dilaton
\begin{equation}
  \label{shet}
  s = \frac{a}2 - \frac{i}2 e^{-\phi} \; ,
\end{equation}
while the rest are given implicitly by
\begin{eqnarray}
  \label{utn}
  A^a_1 &  = & \sqrt2 \frac{n^a - \bar n^a}{u - \bar u} \; , \qquad A^a_2 =
  \sqrt2 \frac{\bar u n^a - u \bar n^a}{u - \bar u} \; , \nn \\
  B_{12} & = & \frac12 \left[(t + \bar t) - \frac{(n + \bar n)^a
      (n- \bar n)^a}{u - \bar u} \right] \; ,\\
  \sqrt G & = & - \frac{i}2
  \left[ (t - \bar t) - \frac{(n- \bar n)^a (n - \bar n)^a}{u - \bar u}
  \right] \; , \nn \\
  G_{11} & = & \frac{2 i}{u - \bar u} \sqrt G \; , \qquad G_{12} = i \frac{u
  + \bar u}{u- \bar u} \sqrt G \; , \nn
\end{eqnarray}
where $A^a_{1,2}$ denote the internal components of the gauge fields, $B_{12}$
is the internal $B$-field, while $G_{11}$, $G_{12}$ and $G$  stand for the
metric on the torus and for its determinant, respectively.

From the $T^2$ compactification point of view, the dynamics of these fields is
naturally described in terms of a $SO(2,n_v - 1)$ matrix $M^{IJ}$ which is
given by
\begin{equation}
  \label{gcfhet}
  M = \left(
    \begin{array}[h]{ccc}
      G^{-1} & - G^{-1} \hat B & - G^{-1} A \\
      -{\hat B}^T G^{-1} &  G + A^T A + {\hat B}^TG^{-1} \hat B & A + {\hat
      B}^T G^{-1} A \\
    -A^T G^{-1} & A^T + A^T G^{-1} \hat B & \mathbf{1}_{n_v-3} + A^T G^{-1} A
      \\
    \end{array}
    \right) \; ,
\end{equation}
where $\hat B_{ij} = B_{ij} + \tfrac12 A^a_i A^a_j$ with indices $i,j$
labeling the $T^2$ directions. The matrix $M$ as defined above leaves
invariant the $SO(2,n_v-1)$ metric
\begin{equation}
  \label{eta}
  \eta = \left(
    \begin{array}[h]{ccc}
      0 & \mathbf{1_2} & 0 \\
      \mathbf{1_2} & 0 & 0 \\
      0 & 0 & \mathbf{1_{n_v-3}} \\
    \end{array} \right)
\end{equation}
in that $M^{IJ} \eta_{JK} M^{KL} = \eta^{IL}$. Then, the kinetic terms of the
moduli are given by
\begin{equation}
  \label{hetkin}
  L_{kin} = \partial_\mu M^{IJ} \partial^\mu (M^{-1})_{IJ} \; ,
\end{equation}
while the gauge kinetic function takes the form
\begin{equation}
  \label{gkfhet}
  I_{IJ} \equiv {\rm Im} \cN_{IJ} = \frac{s - \bar s}{2 i} (M^{-1})_{IJ} \; , \qquad
  R_{IJ} \equiv {\rm Re} \cN_{IJ} = -   \frac{s + \bar s}2 \eta_{IJ} \; .
\end{equation}

The connection to $N=2$ supergravity is not obvious in the above
formulation. Moreover, it turns out that that the natural symplectic basis in
this case is one where no prepotential exists \cite{CCDF,dWKLL,AFGNT}
and so the formulae of appendix
\ref{sg}, and in particular the definition of the gauge coupling matrix
\eqref{Ndef}, do not directly apply. However one can explicitly compute
\eqref{hetkin} using \eqref{gcfhet} and \eqref{utn} and show that these
kinetic terms can be derived from the K\"ahler potential
\begin{equation}
  \label{kpothet}
  K = - \ln \left[i (\bar s- s) \left((u-\bar u) (t - \bar t) - (n-\bar n)^a
  (n- \bar n)^a \right) \right] \; .
\end{equation}
Moreover one can show that using the general formalism of \cite{N=2review} the
gauge coupling matrix \eqref{gkfhet} can be obtained form the following
holomorphic vector
\begin{equation}
  \label{holsec}
  \left(X^I~ |~ F_I \right) = \left(-u,~ 1,~ t,~ ut - n^a n^a,~  \sqrt 2 n^a
      ~|\; -  st,\; -s (ut - n^a n^a),~ su,\; -s,\; - \sqrt2 sn \right)\; ,
\end{equation}
while, obviously, using \eqref{Kspecial} this reproduces the K\"ahler potential
\eqref{kpothet}.
Alternatively, we can start from the type IIA prepotential \eqref{IIAprep}
with the projective coordinates given by
\begin{equation}
  \label{Xid}
  X^0=1 \; , \quad X^1 = s \; , \quad   X^2=u \; , \quad X^3 = t \; , \quad X^a
  = n^a\; .
\end{equation}
Using \eqref{Ndef}, one then computes the gauge coupling matrix $\cN$. To go to
the heterotic symplectic basis \eqref{holsec} we perform the symplectic
rotation with the matrices $U, ~V,~W,~Z$ in \eqref{uvzwg} given by
\begin{eqnarray}
  \label{srot}
  U & = & \left (
    \begin{array}{cccccc}
      0 & 0 & -1 & 0 & 0 \\
      1 & 0 & 0 & 0 & 0 \\
      0 & 0 & 0 & 1 & 0 \\
      0 & 0 & 0 & 0 & 0 \\[.3cm]
      0 & 0 & 0 & 0 & \sqrt2\; \mathbf{1_{n_v-3}} \\
    \end{array}
    \right) , \qquad
    V= \left (
    \begin{array}{cccccc}
      0 & 0 & -1 & 0 & 0 \\
      1 & 0 & 0 & 0 & 0 \\
      0 & 0 & 0 & 1 & 0 \\
      0 & 0 & 0 & 0 & 0 \\[.3cm]
      0 & 0 & 0 & 0 & \tfrac1{\sqrt2} \mathbf{1_{n_v-3}} \\
    \end{array}
    \right) \\[.3cm]
    & & \hspace{2cm}Z = - W = \left (
    \begin{array}{cccccc}
      0 & 0 & 0 & 0 & 0 \\
      0 & 0 & 0 & 0 & 0 \\
      0 & 0 & 0 & 0 & 0 \\
      0 & 1 & 0 & 0 & 0 \\[.3cm]
      0 & 0 & 0 & 0 & \mathbf{0_{n_v-3}} \\
    \end{array}
    \right)\; . \nn
\end{eqnarray}
Note that these matrices are precisely the ones which transform the
holomorphic section derived from the prepotential \eqref{IIAprep} and
\eqref{Xid} into \eqref{holsec}. Moreover, using the transformation of the gauge
coupling matrix \eqref{nchange} it is completely straightforward, but a bit
tedious, to show that the gauge coupling matrix precisely reproduces
\eqref{gkfhet}.
Finally, let us observe that since the matrices $Z$ and $W$ are
non-vanishing this transformation is intrinsically a non-perturbative one in
that it exchanges the gauge field $A^1$ with its magnetic dual, followed by
certain relabelings and rescalings.



\end{document}